\documentclass{article}
\pdfoutput=1
\usepackage{fullpage}
\usepackage{textcomp}
\usepackage{enumerate}
\usepackage{amsmath}
\usepackage{amsfonts}
\usepackage{mathabx}
\usepackage{graphicx}
\usepackage{subfig}
\usepackage{wrapfig}
\usepackage[hidelinks]{hyperref}
\usepackage{algorithm}
\usepackage{algpseudocode}
\usepackage{amsthm}

\newtheorem{lemma}{Lemma}
\newtheorem{theorem}[lemma]{Theorem}
\newtheorem{corollary}[lemma]{Corollary}
\newtheorem{definition}[lemma]{Definition}

\newtheorem{problem}[lemma]{Problem}

\graphicspath{{./figs/}}

\title{On Dominance-Free Samples of a (Colored) Stochastic Dataset}
\author{
    Jie Xue\footnote{Department of Computer Science, University of California, Santa Barbara, CA, USA}\\
    \texttt{xuexx193@umn.edu}
    \and
    Yuan Li\footnote{Facebook Inc., WA, USA}\\
    \texttt{lixx2100@umn.edu}
}
\date{}

\newcommand{\ylOpQuery}{\textsc{Query}}
\newcommand{\ylOpUpdate}{\textsc{Update}}
\newcommand{\ylOpMul}{\textsc{Multiply}}
\newcommand{\ylRT}{\mathcal{T}} % syn function name
\newcommand{\ylGammas}{\varGamma_\mathcal{S}}
\newcommand{\ylRTOneD}{\ylRT_{\text{1D}}}

\begin{document}

\maketitle

\begin{abstract}
A point $p \in \mathbb{R}^d$ is said to \textit{dominate} another point $q \in \mathbb{R}^d$ if the coordinate of $p$ is greater than or equal to the coordinate of $q$ in every dimension.
A set of points in $\mathbb{R}^d$ is \textit{dominance-free} if any two points do not dominate each other.
We consider the problem of counting the dominance-free subsets of a given dataset in $\mathbb{R}^d$, or more generally, computing the probability that a random sample of a stochastic dataset in $\mathbb{R}^d$ where each point is sampled independently with its existence probability is dominance-free.
In fact, we investigate a colored generalization of the problem, in which the points in the given stochastic dataset are colored and we are interested in the random samples that are inter-color dominance-free (i.e., any two points with different colors do not dominate each other).
We propose the first algorithm that solves the problem for $d=2$ in near-quadratic time.
On the other hand, we show that the problem is \#P-hard for any $d \geq 3$, even if the points have a restricted color pattern; this implies the \#P-hardness of the uncolored version (i.e., computing the dominance-free probability of a uncolored stochastic dataset) for $d \geq 3$.
In addition, we show that even when the existence probabilities of the points are all equal to 0.5, the problem remains \#P-hard for any $d \geq 7$; this implies the \#P-hardness of counting dominance-free subsets for $d \geq 7$.
In order to prove our hardness results, we establish some results about embedding the vertices a graph into low-dimensional Euclidean space such that two vertices are connected by an edge in the graph iff they form a dominance pair in the embedding.
These results may be of independent interest and can possibly be applied to other problems.
\end{abstract}

\section{Introduction} \label{sec-intro}
A point $p \in \mathbb{R}^d$ is said to \textit{dominate} another point $q \in \mathbb{R}^d$ (denoted by $p \succ q$) if the coordinate of $p$ is greater than or equal to the coordinate of $q$ in every dimension.
The dominance relation is an important notion in multi-criteria decision-making, and has been well-studied in computational geometry, database, optimization, and other related areas \cite{gabow1984scaling,kung1975finding,papadimitriou2001multiobjective}.
In the last decades, many problems regarding dominance relation have been proposed and investigated, e.g., deciding the dominance-existence of a dataset, counting the number of dominance pairs, reporting the dominance pairs, etc.

In this paper, we consider the following natural question regarding the dominance relation: given a dataset $S$ of points in $\mathbb{R}^d$, how
many subsets of $S$ are \textit{dominance-free} (i.e., any two points in the subset do not dominate each other)?
More generally, one may ask the question in the context of uncertain (or stochastic) datasets: given a \textit{stochastic} dataset $\mathcal{S}$ of points in $\mathbb{R}^d$ where each point is associated with an \textit{existence probability}, what is the probability that a random sample of $\mathcal{S}$ (where each point is sampled independently with its existence probability) is dominance-free?
These questions can be formulated as computation problems of counting dominance-free subsets and computing dominance-free probability.

It is also natural to study a colored generalization of the above problems, in which the given points are \textit{colored}, and we are interested in the subsets or random samples that are \textit{inter-color} dominance-free, i.e., any two points \textit{of different colors} do not dominate each other.
Formally, we formulate the following problems.

\begin{problem}[(Colored) Dominance-free Counting] \label{prob-cdfc}
Given a (colored) dataset $S$ of points in $\mathbb{R}^d$, computing the number of (inter-color) dominance-free subsets of $S$.
\end{problem}

\begin{problem}[(Colored) Dominance-free Probability Computing] \label{prob-cdfpc}
Given a (colored) stochastic dataset $S$ of points in $\mathbb{R}^d$ where each point is associated with an existence probability, computing the probability that a random sample of $\mathcal{S}$ is (inter-color) dominance-free.
\end{problem}

The (Colored) Dominance-free Counting problem defined in Problem~\ref{prob-cdfc} is an instance of counting independent sets in a graph.
Indeed, we can construct a graph whose vertices are points in $S$ and two vertices are connected by an edge if they form a (inter-color) dominance pair, and then the number of (inter-color) dominance-free subsets of $S$ is equal to the number of independent sets in the graph.
Counting independent sets in various types of graphs has been intensively studied in graph theory \cite{cannon2020counting,chandrasekaran2011counting,dyer2002counting,engbers2014counting,samotij2015counting,weitz2006counting,xia20063}.
However, little was known about the problem of counting independent sets in a graph that represents the (inter-color) dominance pairs of a set of (colored) points.

The (Colored) Dominance-free Probability Computing problem defined in Problem~\ref{prob-cdfpc} is a generalization of Problem~\ref{prob-cdfc}.
This problem is an instance of stochastic geometric computing (or geometric computing under uncertainty).
Motivated by dealing with the impreciseness and unreliability of the data obtained in real-world applications due to noise or limitation of devices, a series of work has been done to study geometric problems on stochastic datasets \cite{agarwal2016range,fink2016hyperplane,huang2014epsilon,kamousi2011stochastic2,kamousi2014closest,kumar2016most,xue2016separability,xue2019expected}.
In this topic, a commonly considered problem is to compute the probability that a random sample of a stochastic dataset satisfies some certain property.
Problem~\ref{prob-cdfpc} belongs to this category.

In this paper, we study (Colored) Dominance-free Counting and (Colored) Dominance-free Probability Computing, and give both algorithmic and hardness results for both problems.

\medskip

\noindent
\textbf{Our results.} ($n$ is the number of the points and $d$ is the dimension which is fixed.)
\begin{enumerate}
    \item We solve Problem~\ref{prob-cdfpc} in $O(n^2 \log^2 n)$ time for $d = 2$, showing the both Problem~\ref{prob-cdfc} and~\ref{prob-cdfpc} can be solved in near-quadratic time when $d = 2$.
    \item We prove that, for $d \geq 3$, even the uncolored version of Problem~\ref{prob-cdfpc} is \#P-hard.
    \item We prove that, for $d \geq 7$, even the uncolored version of Problem~\ref{prob-cdfc} is \#P-hard.
    %\item We provide a simple FPRAS for the CSD problem in any dimension.
    %\item We show that the CSD problem is polynomial-time reducible to the FBCSD problem in the same dimension, which implies the \#P-hardness of the latter for $d \geq 3$.
    %\item We solve the FBCSD problem for $d=2$ in $O(n^4 \log^2 n)$ time.
\end{enumerate}

%\medskip

Our algorithm for $d = 2$ uses dynamic programming, with some carefully designed geometric data structure to achieve the desired running time.
To obtain our hardness results, we investigate the problem of embedding the vertices a graph into a low-dimensional Euclidean space such that two vertices are connected by an edge in the graph iff they form a dominance pair in the embedding, which we call a \textit{dominance-preserving embedding} (DPE).
We essentially prove that any 3-regular bipartite planar graph with edge \textit{subdivided} has a DPE into $\mathbb{R}^3$ and any 3-regular bipartite planar graph has a DPE into $\mathbb{R}^7$.
We believe that these results may be of independent interest.

\medskip

\noindent
\textbf{Related work.}
The classical study regarding the dominance relation can be found in many papers such as \cite{gabow1984scaling,kung1975finding,papadimitriou2001multiobjective}.
Recently, there have been a few works considering the dominance relation on stochastic datasets \cite{afshani2011approximate,pei2007probabilistic,zhang2012stochastic};
however, their main focus is the skyline (or dominance-maxima) problems under locational uncertainty, which is quite different from the problems studied in this paper.
Besides problems regarding the dominance relation, many other fundamental geometric problems have also been investigated under stochastic settings in recent years, e.g., closest pair \cite{kamousi2014closest}, minimum spanning tree \cite{kamousi2011stochastic2}, convex hull \cite{agarwal2014convex,suri2013most,xue2019expected}, linear separability \cite{fink2016hyperplane,xue2016separability}, nearest neighbor search \cite{agarwal2013nearest,suri2014most}, range-max query \cite{agarwal2016range}, Voronoi diagrams~\cite{kumar2016most}, coresets \cite{huang2014epsilon}, etc.
Finally, the problem of counting independent sets in various types of graphs has been well-studied~\cite{cannon2020counting,chandrasekaran2011counting,dyer2002counting,engbers2014counting,samotij2015counting,weitz2006counting,xia20063}; see \cite{samotij2015counting} for a survey.
\smallskip

\noindent
\textbf{Basic notions and preliminaries.}
We give the formal definitions of some basic notions used in this paper.
A \textit{colored stochastic dataset} $\mathcal{S}$ in $\mathbb{R}^d$ is a 3-tuple $\mathcal{S} = (S,\text{cl},\pi)$, where $S \subset \mathbb{R}^d$ is the point set, $\text{cl}:S \rightarrow \mathbb{N}$ is the coloring function indicating the colors of the points, and $\pi:S \rightarrow [0,1]$ is the function indicating the existence probabilities of the points, i.e., each point $a \in S$ has the color (label) $\text{cl}(a)$ and the existence probability $\pi(a)$.
A \textit{random sample} of $\mathcal{S}$ refers to a random subset $R \subseteq S$ where each point $a \in S$ is included independently with probability $\pi(a)$.
For any $A \subseteq S$, an \textit{inter-color dominance} in $A$ with respect to $\text{cl}$ (or simply an inter-color dominance in $A$ if the coloring $\text{cl}$ is unambiguous) refers to a pair $(a,b)$ with $a,b \in A$ such that $\text{cl}(a) \neq \text{cl}(b)$ and $a \succ b$.
A \textit{sub-dataset} of $\mathcal{S}$ is a colored stochastic dataset $\mathcal{S}' = (S',\text{cl}',\pi')$ where $S' \subseteq S$, $\text{cl}'=\text{cl}_{|S'}$, $\pi'=\pi_{|S'}$.
A \textit{bipartite} graph is represented as $G = (V \cup V',E)$, where $V,V'$ are the two parts (of vertices) and $E$ is the edge set.
\medskip

\noindent
%\textbf{Paper organization.}
%In Section~\ref{sec-csd}, we investigate the CSD problem.
%In Section~\ref{sec-variant}, we investigate the FBCSD problem.
\textbf{Due to limited space, some proofs are deferred to Appendix~\ref{app-proof}.}
\vspace{-5pt}

\section{The colored stochastic dominance problem} \label{sec-csd}
Let $\mathcal{S} = (S,\text{cl},\pi)$ be a colored stochastic dataset in $\mathbb{R}^d$ with $S = \{a_1,\dots,a_n\}$.
Define $\varGamma_\mathcal{S}$ as the probability that a random sample of $\mathcal{S}$ is inter-color dominance-free.
%Set $\varGamma_\mathcal{S} = 1 - \varLambda_\mathcal{S}$, which is the probability that a random sample of $\mathcal{S}$ contains no inter-color dominances.
The goal of the Colored Dominance-free Probability Computing problem (Problem~\ref{prob-cdfpc}) is to compute $\varGamma_\mathcal{S}$.
%In this section, we consider the problem of computing $\varLambda_\mathcal{S}$ or $\varGamma_\mathcal{S}$, which we call the \textit{colored stochastic dominance} (CSD) problem.

\subsection{An algorithm for \texorpdfstring{$d=2$}{d=2}} \label{sec-2DSD}
The na\"{i}ve method for solving Colored Dominance-free Probability Computing is to enumerate all subsets of $S$ and ``count'' the inter-color dominance-free ones.
However, it requires exponential time, as there are $2^{|S|}$ subsets of $S$ to be considered.
In this section, we show that Colored Dominance-free Probability Computing in $\mathbb{R}^2$ can be solved much more efficiently.
Specifically, we propose an $O(n^2 \log^2 n)$-time algorithm to compute $\varGamma_\mathcal{S}$.
For simplicity, we assume that the points in $S$ have distinct $x$-coordinates and $y$-coordinates; if this is not the case, we can first ``regularize'' $S$ using the following lemma.
Formally, we say a (finite) point set $X \subset \mathbb{R}^d$ is \textit{regular} if $X \subset \{1,2,\dots,|X|\}^d$ and any two distinct points $x,x' \in X$ have distinct coordinates in all dimensions.

\begin{lemma} \label{lem-regular}
    Given a set $S = \{a_1,\dots,a_n\} \subset \mathbb{R}^d$ of distinct points, one can construct in $O(n \log n)$ time a regular set $S_{new} = \{\hat{a}_1,\dots,\hat{a}_n\} \subset \mathbb{R}^d$ such that $\hat{a}_i \succ \hat{a}_j$ iff $a_i \succ a_j$.
\end{lemma}

%Suppose the points $a_1,\dots,a_n \in S$ are already sorted in $<$-order, i.e., $a_1<\cdots<a_n$.
\begin{figure}
    \centering
    \includegraphics[height=3cm]{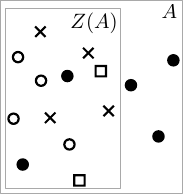}
    \caption{Illustrating $A$ and $Z(A)$.}
    \label{figure:A_and_Z(A)}
\end{figure}
When computing $\varGamma_\mathcal{S}$, we need to consider the random samples which are inter-color dominance-free.
As we will see, in the case of $d=2$, these random samples have good properties, which allows us to solve the problem efficiently in a recursive way.
For any point $a \in \mathbb{R}^2$, we use $x(a)$ (resp., $y(a)$) to denote the $x$-coordinate (resp., $y$-coordinate) of $a$.
Suppose the points $a_1,\dots,a_n \in S$ are already sorted such that $x(a_1) < \cdots< x(a_n)$.
For convenience of exposition, we add a dummy point $a_0$ to $S$ with $x(a_0) < x(a_1)$ and $y(a_0) > y(a_i)$ for all $i \in \{1,\dots,n\}$.
The color $\text{cl}(a_0)$ is defined to be different from $\text{cl}(a_1),\dots,\text{cl}(a_n)$, and $\pi(a_0) = 1$.
Note that including $a_0$ does not change $\varGamma_\mathcal{S}$.
For a subset $A = \{a_{i_1},\dots,a_{i_r}\}$ of $S$ with $i_1<\cdots<i_r$, we define $Z(A) = \emptyset$ if $A$ is monochromatic, and otherwise $Z(A) = \{a_{i_1},\dots,a_{i_l}\}$ such that $\text{cl}(a_{i_l}) \neq \text{cl}(a_{i_{l+1}}) = \cdots = \text{cl}(a_{i_r})$.
In other words, $Z(A)$ is the subset of $A$ obtained by dropping the
    ``rightmost'' points of the same color as $a_{i_r}$;
    see Figure~\ref{figure:A_and_Z(A)}.
We have the following important observation.
%Let $R = \{a_{i_1},\dots,a_{i_r}\}$ be a subset of $S$ where $i_1<\cdots<i_r$.
%We define $Z(R) = \emptyset$ if $R$ is monochromatic, and otherwise $Z(R) = \{a_{i_1},\dots,a_{i_l}\}$ such that $\text{cl}(a_{i_{l+1}}) = \cdots = \text{cl}(a_{i_r}) \neq \text{cl}(a_{i_l})$.
%In other words, $Z(R)$ is the subset of $R$ obtained by removing the ``rightmost'' points of the same color as $a_{i_r}$.
\begin{lemma} \label{lem-struct}
    A random sample $R$ of $\mathcal{S}$ is inter-color dominance-free iff $Z(R)$ is inter-color dominance-free and for any $a \in Z(R)$, $b \in R \backslash Z(R)$ it holds that $y(a) > y(b)$.
\end{lemma}
With this in hand, we then consider how to compute $\varGamma_\mathcal{S}$.
For a nonempty subset $A \subseteq S$, we define the \textit{signature} $\text{sgn}(A)$ of $A$ as a pair $(i,j)$ such that $a_i, a_j \in A$ and $a_i$ (resp., $a_j$) has the greatest $x$-coordinate (resp., smallest $y$-coordinate) among all points in $A$.
Let $E_{i,j}$ be the event that a random sample $R$ of $\mathcal{S}$ is inter-color dominance-free and satisfies $\text{sgn}(R) = (i,j)$.
Note that if a random sample $R$ is inter-color dominance-free, then either $R = \{a_0\}$ or some $E_{i,j}$ happens for $i,j \in \{1,\dots,n\}$.
So we immediately have
\begin{equation*}
    \varGamma_\mathcal{S} = \prod_{i=1}^{n} (1-\pi(a_i)) + \sum_{i = 1}^{n} \sum_{j=1}^{n} \Pr[E_{i,j}].
\end{equation*}
Now the problem is reduced to computing all $\Pr[E_{i,j}]$.
Instead of working on the events $\{E_{i,j}\}$ directly, we consider a set of slightly different events $\{E_{i,j}'\}$ defined as follows.
For $p \in \{0,\dots,n\}$, set $S_p = \{a_0,\dots,a_p\}$, and we use $\mathcal{S}_p$ to denote the sub-dataset of $\mathcal{S}$ with point set $S_p \subseteq S$.
Define $E_{i,j}'$ as the event that a random sample $R$ of $\mathcal{S}_i$ is inter-color dominance-free and satisfies $\text{sgn}(R) = (i,j)$.
It is quite easy to see the equations
\begin{equation*}
    \Pr[E_{i,j}] = \Pr[E_{i,j}'] \cdot \prod_{t=i+1}^{n}(1-\pi(a_t)).
\end{equation*}
Set $F(i,j) = \Pr[E_{i,j}']$.
We show how to compute all $F(i,j)$ recursively by applying Lemma~\ref{lem-struct}.
Observe that $F(i,j) = 0$ if $x(a_i)<x(a_j)$ (equivalently, $i<j$) or $y(a_i) < y(a_j)$ or $\text{cl}(a_i) \neq \text{cl}(a_j)$.
Thus, it suffices to compute all $F(i,j)$ with $i \geq j$, $y(a_i) \geq y(a_j)$, $\text{cl}(a_i) = \text{cl}(a_i)$ (we say the pair $(i,j)$ is \textit{legal} if these three conditions hold).
Let $(i,j)$ be a legal pair.
Trivially, for $i=j=0$, we have $F(i,j) = 1$.
So suppose $i,j>0$.
Let $R$ be a random sample of $\mathcal{S}_i$.
To compute $F(i,j)$, we consider the signature $\text{sgn}(Z(R))$ under the condition that $E_{i,j}'$ happens.
First, when $E_{i,j}'$ happens, we always have $Z(R) \neq \emptyset$, because $R$ at least contains $a_0,a_i,a_j$ (possibly $i = j$) and $\text{cl}(a_0) \neq \text{cl}(a_i) = \text{cl}(a_j)$.
Therefore, in this case, $\text{sgn}(Z(R))$ is defined and must be a legal pair $(i',j')$ for some $i',j' \in \{0,\dots,i-1\}$.
It follows that $F(i,j)$ can be computed by considering for each such pair $(i',j')$ the probability that $R$ is inter-color dominance-free and $\text{sgn}(R) = (i,j)$, $\text{sgn}(Z(R)) = (i',j')$, and then summing up these probabilities.
Note that if $\text{sgn}(R) = (i,j)$ and $\text{sgn}(Z(R)) = (i',j')$, then $i' < j$ and $\text{cl}(i') \neq \text{cl}(i)$.
In addition, if $R$ is inter-color dominance-free, then we must have $y(a_i) < y(a_{j'})$ by Lemma~\ref{lem-struct}.
As such, we only need to consider the legal pairs $(i',j')$ satisfying $i' < j$, $y(a_i) < y(a_{j'})$, $\text{cl}(i') \neq \text{cl}(i)$ (we denote the set of these pairs by $J_{i,j}$).
Fixing such a pair $(i',j') \in J_{i,j}$, we investigate the corresponding probability.
By the definition of $Z(R)$ and Lemma~\ref{lem-struct}, we observe that if $R$ is inter-color dominance-free and $\text{sgn}(R) = (i,j)$, $\text{sgn}(Z(R)) = (i',j')$, then \\
%By addition principle, $F(i,j)$ is just the sum of the probabilities of these two possibilities.
%The possibility of $Z(R) = \emptyset$ occurs only when $R$ is monochromatic.
%In this case, $R$ can never have inter-color dominances.
%So we only need to guarantee that $R$ is monochromatic and $\text{sgn}(R) = (i,j)$.
%In other words, what we need is that $a_i,a_j$ must exist but any points of color different from $\text{cl}(a_i)$ or below $a_j$ must not exist.
%Thus, the probability of this possibility is the product of $\pi_{i,j}^*$ and all $(1-\pi(a_t))$ for $t \in \{1,\dots,i\}$ satisfying $\text{cl}(a_t) \neq \text{cl}(a_i)$ or $y(a_t) < y(a_j)$ (where $\pi_{i,j}^* = \pi(a_i) \cdot \pi(a_j)$ if $i \neq j$ and $\pi_{i,j}^* = \pi(a_i)$ if $i=j$).
%The possibility of $Z(R) \neq \emptyset$ is subtler.
%In this case, we consider the signature $\text{sgn}(Z(R))$.
%Clearly, $\text{sgn}(Z(R))$ must be a legal pair $(i',j')$ for some $i',j' \in \{1,\dots,i\}$.
%By the definition of $Z(R)$, we have that $\text{cl}(i') \neq \text{cl}(i)$.
$\bullet$ $R \cap S_{i'}$ is inter-color dominance-free and $\text{sgn}(R \cap S_{i'}) = (i',j')$; \\
$\bullet$ $R \cap (S_i \backslash S_{i'})$ includes $a_i$ and $a_j$, but does not include any point $a_t$ for $t \in \{i'+1,\dots,i\}$ satisfying $\text{cl}(a_t) \neq \text{cl}(a_i)$ or $y(a_t) < y(a_j)$ or $y(a_{j'}) < y(a_t)$. \\
Conversely, one can also verify that if a random sample $R$ of $\mathcal{S}_i$ satisfies the above two conditions, then $R$ is inter-color dominance-free (by Lemma~\ref{lem-struct}) and $\text{sgn}(R) = (i,j)$, $\text{sgn}(Z(R)) = (i',j')$ (note that $Z(R) = R \cap S_{i'}$).
Therefore, the probability that $R$ is inter-color dominance-free and $\text{sgn}(R) = (i,j)$, $\text{sgn}(Z(R)) = (i',j')$ is just the product $F(i',j') \cdot \pi_{i,j}^* \cdot \varPi_{i,j,i',j'}$, where $\pi_{i,j}^* = \pi(a_i) \cdot \pi(a_j)$ if $i \neq j$ and $\pi_{i,j}^* = \pi(a_i)$ if $i=j$, and $\varPi_{i,j,i',j'}$ is the product of all $(1-\pi(a_t))$ for $t \in \{i'+1,\dots,i\}$ satisfying $\text{cl}(a_t) \neq \text{cl}(a_i)$ or $y(a_t) < y(a_j)$ or $y(a_{j'}) < y(a_t)$.
%By enumerating all such pairs $(i',j')$ and summing up the corresponding probabilities, we can obtain the probability of the possibility of $Z(R) \neq \emptyset$.
%Combining the two possibilities, we have that
Based on this, we can compute $F(i,j)$ as
\begin{equation}
F(i,j) = \sum_{(i',j') \in J_{i,j}} \left( F(i',j') \cdot \pi_{i,j}^* \cdot \varPi_{i,j,i',j'} \right) = \pi_{i,j}^* \cdot \sum_{(i',j') \in J_{i,j}} \left( F(i',j') \cdot \varPi_{i,j,i',j'} \right).
\end{equation}
%where $\pi_{i,j}^* = \pi(a_i) \cdot \pi(a_j)$ if $i \neq j$ and $\pi_{i,j}^* = \pi(a_i)$ if $i=j$, and $\varPi_{i,j,i',j'}$ is the product of all $(1-\pi(a_t))$ for $t \in \{i'+1,\dots,i\}$ satisfying $\text{cl}(a_t) \neq \text{cl}(a_i)$ or $y(a_t) < y(a_j)$ or $y(a_{j'}) < y(a_t)$.
The straightforward way to compute each $F(i,j)$ takes $O(n^3)$ time, which results in an $O(n^5)$-time algorithm for computing $\varGamma_\mathcal{S}$.
Indeed, the runtime of the above algorithm can be drastically improved to
     $O(n^2 \log^2 n)$, by properly using dynamic 2D range trees 
     with some tricks.
We provide some brief ideas here and defer the details to
    Appendix~\ref{appendix:algorithm_via_2d_rangetree}.
    
A legal pair $(i,j)$ is composed of two points, $a_i$ and $a_j$.
We examine all such pairs and compute the corresponding $F(\cdot, \cdot)$ values
    in a fashion by first enumerating $a_i$ from left to right and then
    $a_j$ from bottom to top (when $a_i$ is fixed).
When we are about to compute $F(i,j)$, we need a data structure that stores
    $F(i',j') \cdot \varPi_{i,j,i',j'}$ as the weight for each legal pair
    $(i',j')$ and also supports range sum queries (as only those
    $(i',j') \in J_{i,j}$ are of interest).
A 2D range tree seems to fit here because each legal pair $(i,j)$ can be
    uniquely represented by a 2D point $(x(a_i), y(a_j))$ and 
    we can thus associate on it the corresponding weight.
On the other hand, such a range tree must be dynamic since the weights of
    legal pairs keep varying from time to time.
We show, in Appendix~\ref{appendix:rangetree:details:2016}, how to carefully
    design such a data structure, and more importantly, how to use it to 
    efficiently and correctly update those weights throughout the entire
    computation of $F(i,j)$'s.
As such, we conclude the following.
\begin{theorem}\label{theorem:correct_algorithm}
    The Colored Dominance-free Probability Computing problem for $d=2$ can be solved in $O(n^2 \log^2 n)$ time, so is the Colored Dominance-free Counting problem.
\end{theorem}    

\subsection{Hardness results in higher dimensions}
In this section, we prove the \#P-hardness of Colored Dominance-free Probability Computing (Problem~\ref{prob-cdfpc}) for $d \geq 3$ and Colored Dominance-free Counting (Problem~\ref{prob-cdfc}) for $d \geq 7$.
Indeed, our hardness results are even stronger, which applies to restricted versions of these problems.
For instance, as stated in Section~\ref{sec-intro}, our hardness results covers the uncolored versions.
Furthermore, it also covers the \textit{bichromatic} versions in which the points are colored by two colors.
%We want our hardness result to cover these two specializations.
In order to describe our results, we need to introduce a notion called \textit{color pattern}.

A \textit{partition} of a positive integer $p$ is defined as a multi-set $\varDelta$ of positive integers whose summation is $p$.
In a colored stochastic dataset $\mathcal{S} = (S,\text{cl},\pi)$, the coloring $\text{cl}$ naturally induces a partition of $n = |S|$ given by the multi-set $\{|\text{cl}^{-1}(p)|>0: p \in \mathbb{N}\}$, which we denote by $\varDelta(\mathcal{S})$.
Let $\mathcal{P} = (\varDelta_1,\varDelta_2,\dots)$ be an infinite sequence where $\varDelta_p$ is a partition of $p$.
We say $\mathcal{P}$ is a \textit{color pattern} if it is ``polynomial-time uniform'', i.e., one can compute $\varDelta_p$ for any given $p$ in time polynomial in $p$.
In addition, $\mathcal{P}$ is said to be \textit{balanced} if $p - \max \varDelta_p = \Omega(p^c)$ for some constant $c>0$ (here $\max \varDelta_p$ denotes the maximum in the multi-set $\varDelta_p$).
Then we define the Colored Dominance-free Probability Computing problem with respect to a color pattern $\mathcal{P} = (\varDelta_1,\varDelta_2,\dots)$ as the (standard) Colored Dominance-free Probability Computing problem with the restriction that the input dataset $\mathcal{S} = (S,\text{cl},\pi)$ must satisfy $\varDelta(\mathcal{S}) = \varDelta_n$ where $n = |S|$.

Besides specializing the Colored Dominance-free Probability Computing problem using color pattern, we may also make assumptions for the existence probabilities of the points.
An important case is that all points have the same existence probability equal to $\frac{1}{2}$.
In this case, each of the $2^n$ subsets of $S$ occurs as a random sample of $\mathcal{S}$ with the same probability $2^{-n}$, and computing $\varGamma_\mathcal{S}$ is equivalent to counting the subsets of $S$ satisfying the desired properties.

Our hardness result is presented in the following theorem.
%Naturally, the coloring function $\text{cl}$ induces a partition of $n$ given by the multi-set $\{|\text{cl}^{-1}(p)|>0: p \in \mathbb{N}\}$.
%We will prove the following hardness results.
\begin{theorem} \label{thm-hard}
    Let $\mathcal{P}$ be any balanced color pattern.
    The Colored Dominance-free Probability Computing problem with respect to $\mathcal{P}$ is \#P-hard for $d \geq 3$.
    The Colored Dominance-free Counting problem with respect to $\mathcal{P}$ is \#P-hard for $d \geq 7$.
\end{theorem}
Note that our result above implies the hardness of both the uncolored and bichromatic versions.
The former can be seen via a balanced color pattern $\mathcal{P} = (\varDelta_1,\varDelta_2,\dots)$ with $\varDelta_p = \{1,\dots,1\}$ (i.e., a multi-set consisting of $p$ 1's), while the latter can be seen via a balanced color pattern $\mathcal{P}=(\varDelta_1,\varDelta_2,\dots)$ with $\varDelta_p = \{\frac{p}{2},\frac{p}{2}\}$ for even $p$ and $\varDelta_p = \{\frac{p-1}{2},\frac{p+1}{2}\}$ for odd $p$.
The proof of Theorem~\ref{thm-hard} is nontrivial, so we break it into several stages.

For simplicity, in what follows, we shall identify the Colored Dominance-free Counting problem as a special case of Colored Dominance-free Probability Computing when the existence probabilities of the points are all $\frac{1}{2}$.

\subsubsection{Relation to counting independent sets}
For a colored stochastic dataset $\mathcal{S} = (S,\text{cl},\pi)$, define $G_\mathcal{S} = (S,E_\mathcal{S})$ as the (undirected) graph with vertex set $S$ and edge set $E_\mathcal{S} = \{(a,b): a,b \in S \text{ with } \text{cl}(a) \neq \text{cl}(b) \text{ and } a \succ b\}$.
Since the edges of $G_\mathcal{S}$ one-to-one correspond to the inter-color dominances in $S$, it is clear that a subset $A \subseteq S$ is inter-color dominance-free iff $A$ corresponds to an independent set of $G_\mathcal{S}$.
If $\pi(a) = \frac{1}{2}$ for all $a \in S$, then we immediately have the equation $\varGamma_\mathcal{S} = \mathit{Ind}(G_\mathcal{S})/2^n$, where $\mathit{Ind}(G_\mathcal{S})$ is the number of the independent sets of $G_\mathcal{S}$.
%It follows that, using an oracle that solves the colored stochastic dominance problem in $\mathbb{R}^d$, we are able to count the independent sets of any graph $G \in \mathcal{G}_d$ in polynomial time, where $\mathcal{G}_d$ is any class of graphs for which one .
This observation intuitively tells us the hardness of our problems, as independent-set counting is a well-known \#P-complete problem.
Although we are still far away from proving Theorem~\ref{thm-hard} (because for a given graph $G$ it is not clear how to construct a colored stochastic dataset $\mathcal{S}$ such that $G_\mathcal{S} \cong G$), it is already clear that we should reduce from some independent-set-counting problem.
Regarding independent-set counting, the strongest known result is the following theorem obtained by Xia et al. \cite{xia20063}, which will be used as the origin of our reduction.
\begin{theorem} \label{ind-set-hard}
    Counting independent sets for 3-regular planar bigraphs is \#-P complete.
\end{theorem}
For a graph $G = (V,E)$, we say a map $f:V \rightarrow \mathbb{R}^d$ is a \textit{dominance-preserving embedding} (DPE) of $G$ to $\mathbb{R}^d$ if it satisfies the condition that $(u,v) \in E$ iff $f(u) \succ f(v)$ or $f(v) \succ f(u)$.
We define the \textit{dimension} $\dim(G)$ of $G$ as the smallest number $d$ such that there exists a DPE of $G$ to $\mathbb{R}^d$ (if such a number does not exist, we say $G$ is of infinite dimension).
We have seen above the relation between independent-set counting and Colored Dominance-free Probability Computing with existence probabilities equal to $\frac{1}{2}$ (which is equivalent to Colored Dominance-free Counting). 
%if one can compute a dominance-preserving embedding of $G$ to some $\mathbb{R}^d$, then counting independent sets for $G$ is reduced to an instance of the CSD problem in $\mathbb{R}^d$ with existence probabilities equal to $\frac{1}{2}$.
%when the points in $S$ have existence probabilities $\frac{1}{2}$, computing $\varGamma_\mathcal{S}$ is totally equivalent to counting the independent sets of $G_\mathcal{S}$.
Interestingly, with general existence probabilities, Colored Dominance-free Probability Computing can be related to a much stronger version of independent-set counting, which we call \textit{cardinality-sensitive independent-set counting}.
\begin{definition} \label{def-CSISC}
    Let $c$ be a fixed integer.
    The $c$-\textit{cardinality-sensitive independent-set counting} ($c$-CSISC) problem is defined as follows.
    The input consists of a graph $G = (V,E)$ and a $c$-tuple $\varPhi = (V_1,\dots,V_c)$ of disjoint subsets of $V$.
    The task of the problem is to output, for every $c$-tuple $(n_1,\dots,n_c)$ of integers where $0 \leq n_i \leq |V_i|$, the number of the independent sets $I \subseteq V$ of $G$ satisfying $|I \cap V_i|=n_i$ for all $i \in \{1,\dots,c\}$.
    We denote the desired output by $\mathit{Ind}_\varPhi(G)$, which can be represented by a sequence of $\prod_{i=1}^{c} (|V_i|+1)$ integers.
    Note that the 0-CSISC problem is just the conventional independent-set counting.
\end{definition}
\begin{lemma} \label{lem-CSISC}
    Given any graph $G = (V,E)$ with a DPE $f:V \rightarrow \mathbb{R}^d$ and a $c$-tuple $\varPhi = (V_1,\dots,V_c)$ of disjoint subsets of $V$, one can construct in polynomial time a colored stochastic dataset $\mathcal{S} = (S,\textnormal{cl},\pi)$ in $\mathbb{R}^d$ with $\textnormal{cl}$ injective such that \textnormal{(1)} $G_\mathcal{S} \cong G$ and \textnormal{(2)} $\mathit{Ind}_\varPhi(G)$ can be computed in polynomial time if $\varGamma_\mathcal{S}$ is provided.
    %Furthermore, the construction of $S$ can be done in polynomial time.
    In particular, the $c$-CSISC problem for a class $\mathcal{G}$ of graphs is polynomial-time reducible to Colored Dominance-free Probability Computing in $\mathbb{R}^d$, provided an oracle that computes for any graph in $\mathcal{G}$ a DPE to $\mathbb{R}^d$.
\end{lemma}
\begin{proof}
Suppose $|V| = \{v_1,\dots,v_n\}$.
We construct the colored stochastic dataset $\mathcal{S} = (S,\textnormal{cl},\pi)$ as follows.
Define $S = \{a_1,\dots,a_n\}$ where $a_i = f(v_i) \in \mathbb{R}^d$ and set $\text{cl}(a_i) = i$ (so $\text{cl}$ is injective).
Let $S_1,\dots,S_c$ be the (disjoint) subsets of $S$ corresponding to $V_1,\dots,V_c$ respectively, i.e., $S_i = \{a_j: v_j \in V_i\}$.
Without loss of generality, we may assume $S_1,\dots,S_c$ are all nonempty.
For all points $a \in S_i$, we define $\pi(a) = 4^{-n^{c-i+1}}$ (note that this real number can be represented in polynomial length).
Then for all points $a \in S \backslash (\bigcup_{i=1}^c S_i)$, we define $\pi(a) = \frac{1}{2}$.
With $\mathcal{S}$ constructed above, we already have $G_\mathcal{S} \cong G$, since $f$ is a DPE and all the points in $S$ have distinct colors.
It suffices to show how to ``recover'' $\mathit{Ind}_\varPhi(G)$ from $\varGamma_\mathcal{S}$.
Equivalently, we have to compute, for every $c$-tuple $\phi = (n_1,\dots,n_c)$ of integers where $0 \leq n_i \leq |S_i|$, the number of the subsets $A \subseteq S$ is inter-color dominance-free and satisfying $|A \cap S_i|=n_i$ for all $i \in \{1,\dots,c\}$ (we use $\mathcal{A}_\phi$ to denote the collection of these subsets).
For each $c$-tuple $\phi = (n_1,\dots,n_c)$ with $0 \leq n_i \leq |S_i|$, we notice that any $A \in \mathcal{A}_\phi$ occurs as a random sample of $\mathcal{S}$ with probability
\begin{equation*}
    P_\phi = \frac{1}{2^{n-m}}\prod_{i=1}^{c} \left(\frac{1}{4^{n^{c-i+1}}}\right)^{n_i} \left(1-\frac{1}{4^{n^{c-i+1}}}\right)^{|S_i|-n_i},
\end{equation*}
where $m = \sum_{i=1}^{c}|S_i|$.
Set $N = \prod_{i=1}^{c} (|S_i|+1)$, then we have in total $N$ $c$-tuples $\phi_1,\dots,\phi_N$ (of integers) to be considered ($N$ is polynomial in $n$ as $c$ is constant).
Suppose $\phi_1,\dots,\phi_N$ are already sorted in lexicographical order from small to large.
Our first key observation is that $P_{\phi_i} > 2^n P_{\phi_{i+1}}$ for all $i \in \{1,\dots,N-1\}$.
To see this, assume $\phi_i = (n_1,\dots,n_c)$ and $\phi_{i+1} = (n_1',\dots,n_c')$.
Note that $\phi_1,\dots,\phi_N$ are sorted in lexicographical order, so there exists $k \in \{1,\dots,c\}$ such that $n_j = n_j'$ for all $j<k$ and $n_k' = n_k+1$.
Then it is easy to see that
\begin{equation*}
    \frac{P_{\phi_i}}{P_{\phi_{i+1}}} \geq \frac{1-4^{-n^{c-k+1}}}{4^{-n^{c-k+1}}} \prod_{j=k+1}^{c} (4^{-n^{c-j+1}})^{|S_j|}.
\end{equation*}
If $k=c$, we already have $P_{\phi_i} > 2^n P_{\phi_{i+1}}$.
For the case of $k<c$, since $\sum_{j=k+1}^{c} |S_j| \leq n-1$, the above inequality implies that
\begin{equation*}
    \frac{P_{\phi_i}}{P_{\phi_{i+1}}} \geq \frac{(1-4^{-n^{c-k+1}}) \cdot 4^{-(n-1) \cdot n^{c-k}}}{4^{-n^{c-k+1}}} > 2^n.
\end{equation*}
With this observation in hand, we now consider how to compute $|\mathcal{A}_{\phi_i}|$ for all $i \in \{1,\dots,N\}$ from $\varGamma_\mathcal{S}$.
It is clear that
\begin{equation*}
    \varGamma_\mathcal{S} = \sum_{i=1}^{N} P_{\phi_i} \cdot |\mathcal{A}_{\phi_i}|.
\end{equation*}
For $j \in \{1,\dots,N\}$, we set $\gamma_j = \sum_{i=j+1}^{N} P_{\phi_i} \cdot |\mathcal{A}_{\phi_i}|$.
By the facts that $P_{\phi_i} > 2^n P_{\phi_{i+1}}$ and $\sum_{i=1}^{N} |\mathcal{A}_{\phi_i}| \leq 2^n$, we can deduce $P_{\phi_i} > \gamma_i$ for all $i \in \{1,\dots,N\}$.
Then we are ready to compute $|\mathcal{A}_{\phi_1}|,\dots,|\mathcal{A}_{\phi_N}|$ in order.
Since $P_{\phi_1} > \gamma_1$, $|\mathcal{A}_{\phi_1}|$ must be the greatest integer that is smaller than or equal to $\varGamma_\mathcal{S}/P_{\phi_1}$, and hence can be immediately computed.
Suppose now $|\mathcal{A}_{\phi_1}|,\dots,|\mathcal{A}_{\phi_{m-1}}|$ are already computed, and we consider $|\mathcal{A}_{\phi_m}|$.
Via $|\mathcal{A}_{\phi_1}|,\dots,|\mathcal{A}_{\phi_{m-1}}|$ and $\varGamma_\mathcal{S}$, we may compute $\gamma_{m-1}$.
Because $P_{\phi_m} \geq \gamma_m$, $|\mathcal{A}_{\phi_m}|$ must be the greatest integer that is smaller than or equal to $\gamma_{m-1}/P_{\phi_m}$, and hence can be computed directly.
In this way, we are able to compute all $|\mathcal{A}_{\phi_1}|,\dots,|\mathcal{A}_{\phi_N}|$ and equivalently $\mathit{Ind}_\varPhi(G)$ (in polynomial time).
The statements in the lemma follow readily.
\end{proof}

Indeed, Lemma~\ref{lem-CSISC} has a more general version which reveals the hardness of (general) stochastic geometric problems, see Appendix~\ref{app-general} for details.
Another ingredient to be used in the proof of Theorem~\ref{thm-hard} is a lemma regarding color pattern.
\begin{lemma} \label{lem-color-pattern}
    Let $\mathcal{P} = (\varDelta_1,\varDelta_2,\dots)$ be a balanced color pattern.
    Given a colored stochastic dataset $\mathcal{S} = (S,\textnormal{cl},\pi)$ in $\mathbb{R}^d$ with $\textnormal{cl}$ injective, if $G_\mathcal{S}$ is a bipartite graph, then one can construct in polynomial time another colored stochastic dataset $\mathcal{S}' = (S',\textnormal{cl}',\pi')$ in $\mathbb{R}^d$ satisfying \textnormal{(1)} $\varGamma_{\mathcal{S}'} = \varGamma_\mathcal{S}$, \textnormal{(2)} $S \subseteq S'$, %(meaning that the colors of the points in $S$ may be different from the corresponding ones in $S'$)
    \textnormal{(3)} $\pi'(a) = \frac{1}{2}$ for any $a \in S' \backslash S$,
    \textnormal{(4)} $\langle \mathcal{S}' \rangle$ is an instance of Colored Dominance-free Probability Computing with respect to $\mathcal{P}$.
\end{lemma}

\subsubsection{\#P-hardness for \texorpdfstring{$d \geq 3$}{d >= 3}}
\begin{figure}
    \centering
    \includegraphics[height=2.5cm]{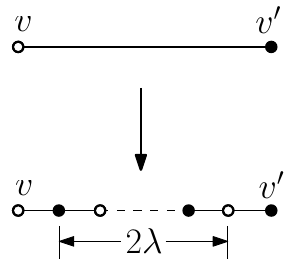}
    \caption{Inserting new vertices to each edge of $G$.}
    \label{fig-2a}
\end{figure}

In this section, we prove the first statement of Theorem~\ref{thm-hard}, by providing a reduction from the independent-set counting problem for 3-regular planar bipartite graphs.
Let $G = (V \cup V',E)$ be a 3-regular planar bipartite graph.
Suppose $|V|=|V'|=n$ (note that we must have $|V|=|V'|$ for $G$ is 3-regular), and then $|E|=3n$.
Instead of working on $G$ directly, we first pass to a new graph, which seems to have a lower dimension.
Set $\lambda = 100n^2$.
We define $G^*$ as the graph obtained from $G$ by inserting $2\lambda$ new vertices to each edge of $G$, i.e., replacing each edge of $G$ with a chain of $2\lambda$ new vertices (see Figure~\ref{fig-2a}).
With an abuse of notation, $V$ and $V'$ are also used to denote the corresponding subsets of the vertices of $G^*$.
Note that $G^*$ is also bipartite, in which $V$ and $V'$ belong to different parts.
We use $U$ (resp., $U'$) to denote the set of the inserted vertices of $G^*$ which belong to the same part as $V$ (resp., $V'$).
Then the two parts of $G^*$ are $V \cup U$ and $V' \cup U'$.
For each edge $e \in E$ of $G$, we denote by $U_e$ (resp. $U_e'$) the set of the $\lambda$ vertices in $U$ (resp., $U'$) which are inserted to the edge $e$.
%Let $\mathcal{G}$ be the class of 3-regular planar bipartite graphs, and $\mathcal{G}^* = \{G^*:G \in \mathcal{G}\}$.

It is not surprising that the independent sets of $G$ are strongly related to those of $G^*$.
%Our first step is to show that the independent sets of $G$ can be counted by solving the 4-CSISC instance $\langle G^*, (V,V',U,U')\rangle$.
Indeed, as we will show, counting independent sets for $G$ can be done by solving the 4-CSISC instance $\langle G^*, (V,V',U,U')\rangle$.
Define $\mathit{Ind}_{p,p'}$ as the number of the independent sets $I$ of $G$ such that $|I \cap V| = p$, $|I \cap V'| = q$.
Also, define $\mathit{Ind}^*_{p,p',q,q'}$ as the number of the independent sets $I^*$ of $G^*$ such that $|I^* \cap V| = p$, $|I^* \cap V'| = p'$, $|I^* \cap U| = q$, $|I^* \cap U'| = q'$.
%We conclude the following.
\begin{lemma} \label{lem-Gstar}
    For any $p,p' \in \{0,\dots,n\}$, we have $\mathit{Ind}_{p,p'} = \mathit{Ind}^*_{p,p',3 \lambda p,3 \lambda n-3 \lambda p}$.
    In particular,
    \begin{equation*}
        \mathit{Ind}(G) = \sum_{i=0}^{n} \sum_{j=0}^{n} \mathit{Ind}_{i,j} = \sum_{i=0}^{n} \sum_{j=0}^{n} \mathit{Ind}^*_{i,j,3 \lambda i,3 \lambda n-3 \lambda i}.
    \end{equation*}
\end{lemma}

Now it suffices to reduce the 4-CSISC instance $\langle G^*, (V,V',U,U')\rangle$ to an instance $\langle \mathcal{S} \rangle$ of Colored Dominance-free Probability Computing in $\mathbb{R}^3$ with respect to a given balanced color pattern $\mathcal{P}$.
Due to Lemma~\ref{lem-CSISC} and \ref{lem-color-pattern}, the only thing we need for the reduction is a DPE of $G^*$ to $\mathbb{R}^3$.
Therefore, our next step is to show $\dim(G^*) \leq 3$ and construct explicitly a DPE of $G^*$ to $\mathbb{R}^3$ (in polynomial time), which is the most non-obvious part of the proof.

Recall that the two parts of $G^*$ are $V \cup U$ and $V' \cup U'$.
The DPE that we are going to construct makes the image of each vertex in $V' \cup U'$ dominates the images of its adjacent vertices in $V \cup U$.
We first consider the embedding for the part $V \cup U$.
Our basic idea is to map the vertices in $V \cup U$ to the plane $H:x+y+z=0$ in $\mathbb{R}^3$.
Note that by doing this we automatically prevent their images from dominating each other.
However, the locations of (the images of) these vertices on $H$ should be carefully chosen so that later we are able to further embed the part $V' \cup U'$ (to $\mathbb{R}^3$) to form a DPE.
Basically, we map $V \cup U$ to $H$ through two steps.
In the first step, the vertices in $V \cup U$ are mapped to $\mathbb{R}^2$ via a map $\varphi:V \cup U \rightarrow \mathbb{R}^2$ to be constructed.
Then in the second step, we properly project $\mathbb{R}^2$ onto $H$ via another map $\psi:\mathbb{R}^2 \rightarrow H$.
By composing $\psi$ and $\varphi$, we obtain the desired map $\psi \circ \varphi: V \cup U \rightarrow H$, which gives us the embedding for $V \cup U$.

\begin{figure}[h]
    \centering
    \includegraphics[height=2.5cm]{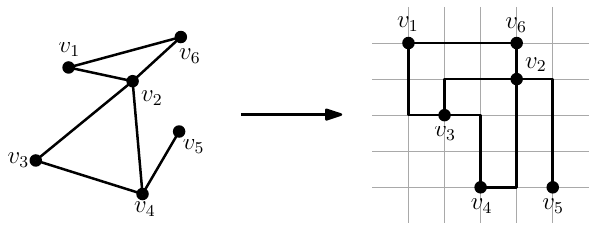}
    \caption{An orthogonal grid drawing.}
    \label{fig-2b}
\end{figure}

\begin{figure}[htpb]
    \centering
    \subfloat[]{
        \includegraphics[height=2.5cm]{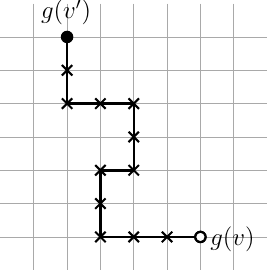}
        \label{figure:2c-1}
    }
    \hspace{1cm}
    \subfloat[]{
        \includegraphics[height=2.5cm]{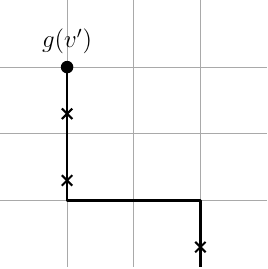}
        \label{figure:2c-2}
    }
    \hspace{1cm}
    \subfloat[]{
        \includegraphics[height=2.5cm]{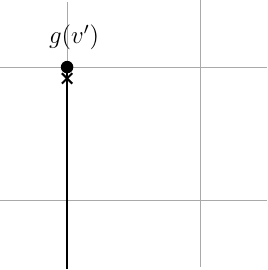}
        \label{figure:2c-3}
    }
    \caption{The construction of $P_e$.}
\end{figure}

To construct $\varphi$, we need a notion about graph drawing.
Let $K = (\mathbb{Z} \times \mathbb{R}) \cup (\mathbb{R} \times \mathbb{Z}) \subset \mathbb{R}^2$ be the grid.
An \textit{orthogonal grid drawing} (OGD) of a (planar) graph is a planar drawing with image in the grid $K$ such that the vertices are mapped to the grid points $\mathbb{Z}^2$.
Note that an OGD draws the edges of the graph as (non-intersecting) orthogonal curves in $\mathbb{R}^2$ consisting of unit-length horizontal/vertical segments each of which connects two adjacent grid points (see Figure~\ref{fig-2b}).
We will apply the following result from \cite{valiant1981universality}.
\begin{theorem}
    For any $t$-vertex planar graph of (maximum) degree 3, one can compute in polynomial time an OGD with image in $K \cap Q_{3t}$ where $Q_i$ denotes the square $[1,i] \times [1,i] \subset \mathbb{R}^2$.
\end{theorem}
Consider the original 3-regular planar bipartite graph $G = (V \cup V',E)$.
By applying the above theorem, we can find an OGD $g$ for $G$ with image in $K \cap Q_{6n}$.
For each vertex $v \in V \cup V'$ of $G$, we denote by $g(v)$ the image of $v$ in $\mathbb{R}^2$ under the OGD $g$.
Also, for each edge $e = (v,v') \in E$ of $G$, we denote by $g(e)$ the image of $e$ under $g$, which is an orthogonal curve in $\mathbb{R}^2$ connecting $g(v)$ and $g(v')$.
With the OGD $g$ in hand, we construct the map $\varphi$ as follows.
For all $v \in V$, we simply define $\varphi(v) = g(v)$.
To determine $\varphi(u)$ for $u \in U$, we consider the vertices in $U_e$ for each edge $e \in E$ of $G$ separately.
Suppose $e = (v,v')$ and $U_e = \{u_1,\dots,u_\lambda\}$ where $u_1,\dots,u_\lambda$ are sorted in the order they appear on $e$ (from $v$ to $v'$).
%Let $\hat{e}$ be the image of $e$ under the OGD, which is an orthogonal curve connecting $\varphi(v)$ and $\varphi(v')$.
Consider the curve $g(e)$.
Since $g$ is an OGD, $g(e)$ must consist of unit-length horizontal/vertical segments (each of which connects two grid points).
The total number $m$ of these unit segments is upper bounded by $(6n)^2$ as $g(e) \subset K \cap Q_{6n}$.
Now we pick a set $P_e$ of $\lambda$ (distinct) points on $g(e)$ as follows. \\
$\bullet$ The $m-1$ grid points in the interior of $g(e)$ are included in $P_e$ (see Figure~\ref{figure:2c-1}). \\
$\bullet$ On each unit vertical segment of $g(e)$, we pick the point with distance $0.3$ from the bottom endpoint and include it to $P_e$ (see Figure~\ref{figure:2c-2}). \\
$\bullet$ On the unit segment of $g(e)$ adjacent to $g(v')$, we pick the point with distance $0.01$ from $g(v')$ and include it to $P_e$ (see Figure~\ref{figure:2c-3}). \\
$\bullet$ Note that the number of the above three types of points is at most $2m \leq 72n^2 < \lambda$.
To make $|P_e| = \lambda$, we then arbitrarily pick more (distinct) points on $g(e)$ which have distances at least $0.4$ to any grid point, and add them to $P_e$. \\
Suppose $P_e = \{r_1,\dots,r_\lambda\}$ where $r_1,\dots,r_\lambda$ are sorted in the order they appear on the curve $g(e)$ (from $g(v)$ to $g(v')$).
We then define $\varphi(u_i) = r_i$.
We do the same thing for every edge $e \in E$ of $G$.
In this way, we determine $\varphi(u)$ for all $u \in U$ and complete defining the map $\varphi$.
%Now we have mapped $V \cup U$ to $\mathbb{R}^2$.
The next step, as mentioned before, is to project $\mathbb{R}^2$ onto $H$.
The projection map $\psi:\mathbb{R}^2 \rightarrow H$ is defined as $\psi:(x,y) \mapsto (x+y,y-x,-2y)$.
Then the composition $\psi \circ \varphi: V \cup U \rightarrow H$ gives us the first part of our DPE.
The remaining task is to embed the part $V' \cup U'$ to $\mathbb{R}^3$, which completes the construction of our DPE.
We must guarantee that the image of each vertex $w' \in V' \cup U'$ dominates and only dominates the images of the vertices in $V \cup U$ adjacent to $w'$.
To achieve this, we first establish an important property of the map $\psi \circ \varphi: V \cup U \rightarrow H$ constructed above.
%Before doing this, we first explain why we map $V \cup U$ to $H$ in the above way.
For a finite set $A$ of points in $\mathbb{R}^d$, we define a point $\underline{\max}(A) \in \mathbb{R}^d$ as the \textit{coordinate-wise} maximum of $A$, i.e., the $i$-th coordinate of $\underline{\max}(A)$ is the maximum of the $i$-th coordinates of all points in $A$, for all $i \in \{1,\dots,d\}$.
%Our construction above satisfies the following.
\begin{lemma} \label{lem-prop}
    For each vertex $w' \in V' \cup U'$, let $\textnormal{Adj}_{w'} \subseteq V \cup U$ be the set of the vertices adjacent to $w'$ in $G^*$, and $A_{w'} = (\psi \circ \varphi)(\textnormal{Adj}_{w'}) \subset \mathbb{R}^3$ be the set of the corresponding images under $\psi \circ \varphi$.
    Then for any $w \in V \cup U$ and $w' \in V' \cup U'$, the point $\underline{\max}(A_{w'}) \in \mathbb{R}^3$ dominates $(\psi \circ \varphi)(w)$ iff $w \in \textnormal{Adj}_{w'}$.
\end{lemma}
\begin{proof}
The ``if'' part is obvious, because $\underline{\max}(A)$ clearly dominates every point in $A$ for any (finite) $A \subset \mathbb{R}^d$ with $|A| \geq 2$ (note that $|A_{w'}| \geq 2$ for any $w' \in V' \cup U'$).
It suffices to prove the ``only if'' part.
For a point $p \in \mathbb{R}^3$, we denote by $H_p$ the set of the points on the plane $H$ which are dominated by $p$.
We first observe that if $H_p \neq \emptyset$, then the preimage $\psi^{-1}(H_p)$ of $H_p$ under $\psi$ (which is a region in $\mathbb{R}^2$) must be a (closed) right-angled isosceles triangle in $\mathbb{R}^2$ whose hypotenuse is horizontal (we call this kind of triangles \textit{standard} triangles).
To see this, assume $p = (x_p,y_p,z_p)$ and $H_p \neq \emptyset$ (this is equivalent to saying $x_p+y_p+z_p>0$).
Then $\psi^{-1}(H_p)$ consists of all the points $(x,y) \in \mathbb{R}^2$ satisfying $x+y \leq x_p$, $y-x \leq y_p$, $y \geq -z_p/2$, and hence is a standard triangle.
Furthermore, it is easy to see that if $p = \underline{\max}(A)$ for a finite set $A \subset H$ with $|A| \geq 2$, then $H_p \neq \emptyset$ and $\psi^{-1}(H_p)$ is the minimal standard triangle containing $\psi^{-1}(A)$ (by ``minimal'' we mean that any standard triangle containing $\psi^{-1}(A)$ is a superset of $\psi^{-1}(H_p)$, both as subsets of $\mathbb{R}^2$, see Figure~\ref{fig-Aa}).
Therefore, we only need to show that for any vertex $w' \in V' \cup U'$, the minimal standard triangle containing $\psi^{-1}(A_{w'}) = \varphi(\textnormal{Adj}_{w'})$ does not contain $\varphi(w)$ for any vertex $w \in (V \cup U) \backslash \textnormal{Adj}_{w'}$.
We consider two cases, $w' \in V'$ and $w' \in U'$.
\begin{figure}[h]
    \centering
    \includegraphics[height=2.5cm]{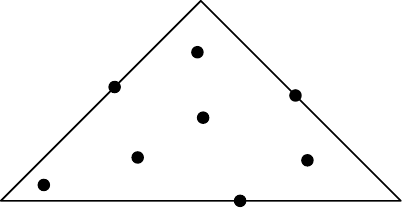}
    \caption{The minimal standard triangle in $\mathbb{R}^2$ containing a set of points.}
    \label{fig-Aa}
\end{figure}
\begin{figure}[h]
    \centering
    \includegraphics[height=3cm]{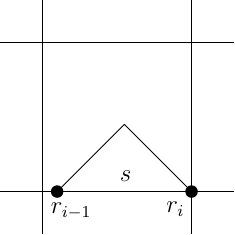}
    \caption{The case that $s$ is horizontal.}
    \label{fig-Ab}
\end{figure}
\begin{figure}[h]
    \centering
    \includegraphics[height=2.5cm]{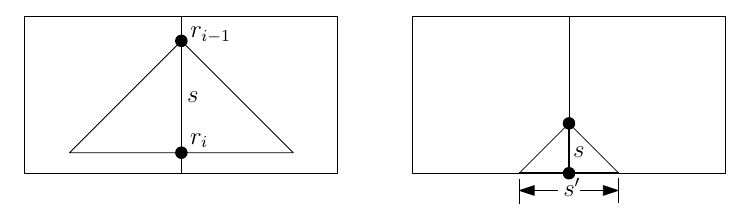}
    \caption{The case that $s$ is vertical.}
    \label{fig-Ac}
\end{figure}

In the case $w' \in V'$, $\textnormal{Adj}_{w'}$ consists of three vertices (for $G$ is 3-regular) in $U$, say $w_1,w_2,w_3$.
Recall that $g$ is the OGD of $G$ used in constructing the map $\varphi$.
By our construction of $\varphi$, we see that each of $\varphi(w_1),\varphi(w_2),\varphi(w_3)$ has distance 0.01 from $g(w')$.
On the other hand, one can easily verify that for any vertex $w \in (V \cup U) \backslash \textnormal{Adj}_{w'}$, $\varphi(w)$ is ``far away'' from $g(w')$ (more precisely, with distance at least 0.3).
Therefore, the minimal standard triangle containing $\varphi(w_1),\varphi(w_2),\varphi(w_3)$ does not contain $\varphi(w)$ for any vertex $w \in (V \cup U) \backslash \textnormal{Adj}_{w'}$.

In the case of $w' \in U'$, we may assume $w' \in U_e'$ for some edge $e = (v,v') \in E$ of $G$.
Then $\textnormal{Adj}_{w'}$ consists of two vertices in $\{v\} \cup U_e$, say $w_1,w_2$.
Recall that $P_e$ is the set of the $\lambda$ points chosen on the curve $g(e)$ for sake of defining $\varphi(u)$ for $u \in U_e$.
As before, we suppose $P_e = \{r_1,\dots,r_\lambda\}$ where $r_1,\dots,r_\lambda$ are sorted in the order they appear on the curve $g(e)$ (from $g(v)$ to $g(v')$).
For convenience, set $r_0 = g(v)$.
Then we may assume $\varphi(w_1) = r_{i-1}$ and $\varphi(w_2) = r_i$ for some $i \in \{1,\dots,\lambda\}$.
Let $s = \overline{r_{i-1} r_i}$ be the segment in $\mathbb{R}^2$ with endpoints $r_{i-1}$ and $r_i$, and $\triangle$ be the minimal standard triangle containing $r_{i-1}$ and $r_i$.
Since all the grid points on $g(e)$ are included in $P_e$, $s$ must be a horizontal or vertical segment contained in $g(e)$.
Furthermore, the interior of $s$ does not contain $\varphi(w)$ for any vertex $w \in V \cup U$ and in particular does not contain any grid points.
We discuss two cases separately: $s$ is horizontal and $s$ is vertical.
Recall that $K = (\mathbb{Z} \times \mathbb{R}) \cup (\mathbb{R} \times \mathbb{Z}) \subset \mathbb{R}^2$ is the grid.
If $s$ is horizontal, then $\triangle$ is just the standard triangle having $s$ as its hypotenuse (see Figure~\ref{fig-Ab}).
In this case, we have $\triangle \cap K = s$, which implies that $\triangle$ does not contain $\varphi(w)$ for any vertex $w \in (V \cup U) \backslash \{w_1,w_2\}$.
For the case that $s$ is vertical, assume that $r_{i-1}$ is the top endpoint and $r_i$ is the bottom one.
Then $r_{i-1}$ is the right-angled vertex of $\triangle$, and $r_i$ is the midpoint of the hypotenuse of $\triangle$.
If $r_i$ is not a grid point, we again have $\triangle \cap K = s$ and thus we are done (see the left part of Figure~\ref{fig-Ac}).
If $r_i$ is a grid point, the distance between $r_{i-1}$ and $r_i$ must be 0.3, by our construction of $P_e$.
In this situation, $\triangle \cap K$ consists of $s$ and a horizontal segment $s'$ of length 0.6 which is the hypotenuse of $\triangle$ (see the right part of Figure~\ref{fig-Ac}).
We claim that $\varphi(w)$ is not on $s'$ for any vertex $w \in (V \cup U) \backslash \{w_2\}$.
Indeed, by our construction of $\varphi$, if $\varphi(w)$ is in the interior of some unit horizontal segment, then $\varphi(w)$ is either with distance 0.01 from $g(v')$ for some $v' \in V'$ or with distance at least 0.4 from any grid point.
In each of the cases, $\varphi(w)$ is ``far away'' from $r_i$ (more precisely, with distance at least 0.4).
But any point on $s'$ has distance at most 0.3 from $r_i$.
Therefore, $\varphi(w)$ is not on $s'$.
It immediately follows that $\triangle$ does not contain $\varphi(w)$ for any vertex $w \in (V \cup U) \backslash \{w_1,w_2\}$, which completes the proof.
\end{proof}

Once the above property is revealed, the construction of the map $V' \cup U' \rightarrow \mathbb{R}^3$ is quite simple: we just map each vertex $w' \in V' \cup U'$ to the point $\underline{\max}(A_{w'}) \in \mathbb{R}^3$.
Now we complete constructing the embedding of $G^*$ to $\mathbb{R}^3$, and need to verify it is truly a DPE.
%We simply map each vertex $w' \in V' \cup U'$ to the point $\underline{\max}(A_{w'})$.
Lemma~\ref{lem-prop} already guarantees that the image of each $w' \in V' \cup U'$ dominates (the images of) the vertices in $\text{Adj}_{w'}$ (i.e., the vertices in $V \cup U$ that are adjacent to $w'$) but does not dominate (the images of) any other vertices in $V \cup U$.
%This is what we desire since $A_{w'}$ is the set of the vertices adjacent to $w'$.
So it suffices to show that the images of the vertices in $V' \cup U'$ do not dominate each other.
Let $w_1', w_2'\in V' \cup U'$ be two distinct vertices, and assume that $\underline{\max}(A_{w_1'})$ dominates $\underline{\max}(A_{w_2'})$.
Then we must have $\underline{\max}(A_{w_1'})$ dominates the points in $A_{w_2'}$.
By Lemma~\ref{lem-prop}, this implies that $\text{Adj}_{w_2'} \subseteq \text{Adj}_{w_1'}$.
However, as one can easily see from the structure of $G^*$, it never happens that $\text{Adj}_{w_2'} \subseteq \text{Adj}_{w_1'}$ unless $w_1' = w_2'$.
Thus, we conclude that the map constructed is a DPE of $G^*$ to $\mathbb{R}^3$.
With the DPE in hand, by applying Lemma~\ref{lem-CSISC} and \ref{lem-color-pattern}, the first statement of Theorem~\ref{thm-hard} is readily proved.
%The remaining points in $P_e$ are chosen from $\hat{e}$ such that they have distances at least $0.4$ to any grid point.
%First, the $(m-1)$ grid points on $\hat{e}$ are included in $P_e$.
%Second, for each vertical unit segment of $\hat{e}$, we add to $P_e$ the point on the segment with distance $0.3$ from the bottom grid point (and thus $0.7$ from the top grid point).

\subsubsection{\#P-hardness for \texorpdfstring{$d \geq 7$}{d >= 7} with $\frac{1}{2}$ existence probabilities} \label{sec-d>=7}
In this section, we prove the second statement of Theorem~\ref{thm-hard}.
When the existence probabilities are restricted to be $\frac{1}{2}$, we are no longer able to apply the tricks used in the previous section, as the reduction from the CSISC problem (Lemma~\ref{lem-CSISC}) cannot be done under such a restriction.
This is the reason for why we have to ``loosen'' the dimension to 7 in this case.
%As we have seen, when the existence probabilities are restricted to be $\frac{1}{2}$, computing $\varGamma_\mathcal{S}$ is equivalent to counting independent sets of $G_\mathcal{S}$.
%That is why we have to ``loosen'' the dimension from 3 to 7.

As we have seen, for a colored stochastic dataset $\mathcal{S} = (S,\text{cl},\pi)$ with $\pi(a) = \frac{1}{2}$ for all $a \in S$, computing $\varGamma_\mathcal{S}$ is totally equivalent to counting independent sets for $G_\mathcal{S}$.
Therefore, we complete the proof by establishing a more direct reduction from independent-set counting for 3-regular planar bipartite graphs, which constructs directly a DPE of the input graph to $\mathbb{R}^7$.
However, it is non-obvious that any 3-regular planar bipartite graph $G$ has dimension at most 7 and how to construct a DPE of $G$ to $\mathbb{R}^7$ in polynomial time.
To prove this, we introduce a new technique based on graph coloring.
Indeed, we consider a more general case in which the graph $G$ is an arbitrary bipartite graph.
The graph coloring to be used is slightly different from the conventional notion, which we call \textit{halfcoloring}.
Let $G=(V \cup V',E)$ be a bipartite graph.
For any two distinct vertices $u,v \in V$, we define $u \sim v$ if there exists a vertex in $V'$ adjacent to both $u$ and $v$.
%(note that $u \sim v$ only if they belong to the same part of $G$)
\begin{definition}
    A $k$-\textit{halfcoloring} of $G$ on $V$ is a map $h:V \rightarrow \{1,\dots,k\}$.
    The halfcoloring $h$ is said to be \textit{discrete} if $h(u) \neq h(v)$ for any $u,v \in V$ with $u \sim v$, to be \textit{semi-discrete} if it satisfies the condition that for any distinct $u,v,w \in V$ with $u \sim v$ and $v \sim w$, $h(u), h(v), h(w)$ are not all the same.
    Symmetrically, we may also define halfcoloring on $V'$.
\end{definition}
We may relate halfcoloring to the conventional graph coloring as follows.
Define $G' = (V,E')$ with $E' = \{(u,v): u \sim v \text{ in }G\}$.
Clearly, a discrete $k$-halfcoloring of $G$ on $V$ corresponds to a (conventional) $k$-coloring of $G'$ satisfying that no two adjacent vertices share the same color, i.e., the subgraph of $G'$ induced by each color form an independent set of $G'$.
Similarly, a semi-discrete $k$-halfcoloring of $G$ on $V$ corresponds to a $k$-coloring of $G'$ satisfying that the subgraph of $G'$ induced by each color consists of connected components of sizes at most 2.
If $h$ is a $k$-halfcoloring of $G$ on $V$, then for each $v' \in V'$ we denote by $\chi_h(v')$ the number of the colors ``adjacent'' to $v'$ (the color $i$ is said to be adjacent to $v'$ if there is a vertex $v \in V$ adjacent to $v'$ with $h(v) = i$).
Our technical result is the following theorem, which establishes a relation between halfcoloring and graph dimension.
\begin{theorem} \label{thm-embedding}
    Let $G=(V \cup V',E)$ be a bipartite graph.
    If there exists a semi-discrete $k$-halfcoloring $h:V \rightarrow \{1,\dots,k\}$ of $G$ (on $V$), then $\dim(G) \leq 2k$.
    In addition, if $\chi_h(v') < k$ for all  $v' \in V'$, then $\dim(G) \leq 2k-1$.
    Furthermore, with $h$ in hand, one can compute in polynomial time a DPE of $G$ to $\mathbb{R}^{2k}$, or $\mathbb{R}^{2k-1}$ in the latter case.
\end{theorem}

We then apply the halfcoloring technique to show that $\dim(G) \leq 7$ for any 3-regular planar bipartite graph $G$, which will give us a proof for the second statement of Theorem~\ref{thm-hard}.
To achieve this, the only missing piece is the following observation.
\begin{lemma} \label{lem-halfcoloring}
    Every 3-regular planar bipartite graph has a discrete 4-halfcoloring, which can be computed in polynomial time.
\end{lemma}
Now it is quite straightforward to prove the second statement of Theorem~\ref{thm-hard}.
Let $G$ be a 3-regular planar bipartite graph.
By combining Theorem~\ref{thm-embedding} and Lemma~\ref{lem-halfcoloring}, we can compute a DPE of $G$ to $\mathbb{R}^7$ in polynomial time.
By taking the images of the vertices of $G$ under the DPE, we obtain a set $S$ of points in $\mathbb{R}^7$.
Using the point set $S$, we further construct a colored stochastic dataset $\mathcal{S} = (S,\text{cl},\pi)$ by choosing an injection $\text{cl}:S \rightarrow \mathbb{N}$ and defining $\pi(a) = \frac{1}{2}$ for any $a \in S$.
It is clear that $G_\mathcal{S} \cong G$ and thus $\mathit{Ind}(G) = 2^{|S|} \varGamma_\mathcal{S}$.
Then by applying Lemma~\ref{lem-color-pattern}, we can compute another colored stochastic dataset $\mathcal{S}' = (S',\text{cl}',\pi')$ such that $\varGamma_{\mathcal{S}'} = \varGamma_\mathcal{S}$ and $\pi'(a) = \frac{1}{2}$ for any $a \in S'$, and more importantly, $\langle \mathcal{S}' \rangle$ is an instance of Colored Dominance-free Probability Computing with respect to $\mathcal{P}$.
Since Colored Dominance-free Counting is a special case of Colored Dominance-free Probability Computing when the points have existence probabilities equal to $\frac{1}{2}$, the second statement of Theorem~\ref{thm-hard} is proved.
Our conclusion that any 3-regular planar bipartite graph has dimension at most 7 is of independent interest, which has an implication in order dimension theory \cite{trotter2001combinatorics} (see Appendix~\ref{app-order}).

\bibliography{my_bib}

\newpage

\appendix{\noindent \huge \textsf{Appendix}}
\section{Missing proofs} \label{app-proof}

\subsection{Proof of Lemma~\ref{lem-regular}}
Fixing $p \in \{1,\dots,d\}$, we determine the $p$-th coordinates of $\hat{a}_1,\dots,\hat{a}_n$ as follows.
For all $i \in \{1,\dots,n\}$, define a triple $\phi_i=(\gamma_i,\sigma_i,i)$ where $\gamma_i$ is the $p$-th coordinate of $a_i$ and $\sigma_i$ is the sum of the $d$ coordinates of $a_i$.
Then we sort all $\phi_i$ in lexicographic order from small to large, and suppose $\phi_{i_1},\dots,\phi_{i_n}$ is the resulting sorted sequence.
We have $\phi_{i_1}<\cdots<\phi_{i_n}$ under lexicographic order, since there exist no ties.
Now we simply set the $p$-th coordinates of $\hat{a}_{i_1},\dots,\hat{a}_{i_n}$ to be $1,\dots,n$ respectively.
In this way, we obtain the new set $S_{new} = \{\hat{a}_1,\dots,\hat{a}_n\} \subset \mathbb{R}^d$ in $O(n \log n)$ time (note that $d$ is assumed to be constant).
It is clear that $S_{new}$ is regular.
We verify that $S_{new}$ satisfies the desired property.
Assume $a_i \succ a_j$.
Then in each dimension, the coordinate of $a_i$ is greater than or equal to the coordinate of $a_j$.
In addition, the sum of the $d$ coordinates of $a_i$ is greater than that of $a_j$.
Therefore, in all dimensions, the coordinates of $\hat{a}_i$ are greater than the coordinates of $\hat{a}_j$, i.e., $\hat{a}_i \succ \hat{a}_j$.
Assume $a_i \nsucc a_j$.
Then there exists $p \in \{1,\dots,d\}$ such that the $p$-th coordinate of $a_i$ is smaller than the $p$-th coordinate of $a_j$.
By definition, the $p$-th coordinate of $\hat{a}_i$ is also smaller than the $p$-th coordinate of $\hat{a}_j$.
Therefore, $\hat{a}_i \nsucc \hat{a}_j$.

\subsection{Proof of Lemma~\ref{lem-struct}}
To see the ``if'' part, assume that $Z(R)$ is inter-color dominance-free and $y(a) > y(b)$ for any $a \in Z(R)$, $b \in R \backslash Z(R)$.
In this case, any two points in $Z(R)$ cannot form an inter-color dominance.
Also, any two points in $R \backslash Z(R)$ cannot form an inter-color dominance for $R \backslash Z(R)$ is monochromatic.
It suffices to show that any $a \in Z(R)$ and $b \in R \backslash Z(R)$ cannot form an inter-color dominance.
By assumption, we have $y(a) > y(b)$.
But by the definition of $Z(S)$, we also have $x(a) < x(b)$.
Thus, $a$ and $b$ do not dominate each other.
To see the ``only if'' part, assume $R$ is inter-color dominance-free.
Since $Z(R)$ is a subset of $R$, it is also inter-color dominance-free.
Let $a \in Z(R)$ and $b \in R \backslash Z(R)$ be two points.
As argued before, we have $x(a) < x(b)$.
If $\text{cl}(a) \neq \text{cl}(b)$, then it is clear that $y(a) > y(b)$ (otherwise $(a,b)$ forms an inter-color dominance).
The only remaining case is $\text{cl}(a) = \text{cl}(b)$.
Since $a \in Z(R)$, by the definition of $Z(R)$, we may find a point $o \in Z(R)$ such that $x(a) < x(o) < x(b)$ and $\text{cl}(o) \neq \text{cl}(a) = \text{cl}(b)$.
If $y(a) < y(b)$, then either $y(a) < y(o)$ or $y(o) < y(b)$, i.e., either $(a,o)$ or $(o,b)$ forms an inter-color dominance.
Because $R$ is inter-color dominance-free, we must have $y(a) > y(b)$.

\subsection{Proof of Lemma~\ref{lem-color-pattern}}
Since $\mathcal{P}$ is balanced, we can find an constants $c>0$ such that $n- \max \varDelta_n \geq n^c$ for any sufficiently large $n$.
Suppose $G_\mathcal{S} = (V \cup V',E)$ where $|V|=n$ and $|V'|=n'$.
We may write $S = \{a_1,\dots,a_{n+n'}\}$ where $a_1,\dots,a_n$ correspond to the vertices in $V$ and $a_{n+1},\dots,a_{n+n'}$ correspond to those in $V'$.
Because $\text{cl}$ is injective (i.e., the points in $S$ are of distinct colors), we have that $a_1,\dots,a_n$ do not dominate each other, and the same holds for $a_{n+1},\dots,a_{n+n'}$.
Set $N = \max\{2n+n',(n')^{1/c}\}$.
Now we construct $\mathcal{S}' = (S',\text{cl}',\pi')$ as follows.
First, we pick a set $A$ of $N-(n+n')$ points in $\mathbb{R}^d$ which do not dominate each other and do not form dominances with any points in $S$.
Set $S' = S \cup A$, so $S \subseteq S'$ and $|S'| = N$.
The points in $A$ are used as dummy points, and can never influences $\varGamma_{\mathcal{S}'}$ (since they are not involved in any dominances).
With a little bit abuse of notation, we also use $a_1,\dots,a_{n+n'}$ to denote the non-dummy points in $S'$.
We then define $\pi'$ as $\pi'(a) = \pi(a)$ for $a \in S$ and $\pi'(a) = \frac{1}{2}$ for $a \in A$.
It suffices to assign colors to the points in $S'$, i.e., define the coloring function $\text{cl}'$.
% $S' = S \cup A$ of $N$ stochastic points by adding $N-(n+n')$ dummy points to $S$ with existence probabilities equal to $\frac{1}{2}$ (ignoring the colors of the points for a while).
%The coordinates of the dummy points added are carefully chosen in $\mathbb{R}^d$ such that each of them does not dominate or be dominated by any other points in $S'$.
%Then we assign colors to the points in $S'$.
Since we want $\langle \mathcal{S}' \rangle$ to be an instance of Colored Dominance-Free Probability Computing with respect to $\mathcal{P}$, the coloring $\text{cl}'$ must induce the partition $\varDelta_N$ of $N$.
Suppose $\varDelta_N = \{r_1,\dots,r_k\}$ (as a multi-set) where $r_1 \geq \cdots \geq r_k$.
%We try to construct $\text{cl}'$ such that the image of $\text{cl}'$ is $\{1,\dots,k\}$ and the preimage of each $i$ under $\text{cl}'$ has size $r_i$ (i.e., there are exactly $r_i$ points in $S'$ painted with color $i$ by $\text{cl}'$) for $i \in \{1,\dots,k\}$.
Let $l$ be the smallest integer such that $\sum_{i=1}^{l} r_i \geq n$.
It is easy to see that $\sum_{i=l+1}^{k} r_i \geq n'$.
Indeed, if $l=1$, then we have 
\begin{equation*}
    \sum_{i=2}^{m} r_i = N - \max \varDelta_N \geq N^c \geq n'
\end{equation*}
by assumption.
In the case of $l>1$, we have that $\sum_{i=1}^{l} r_i < 2n$ and thus $\sum_{i=l+1}^{k} r_i > N-2n \geq n'$.
This fact implies that we are able to define the coloring function $\text{cl}'$ with image $\{1,\dots,k\}$ such that (1) there are exactly $r_i$ points in $S'$ mapped to the color $i$ by $\text{cl}'$, (2) $\text{cl}'(a) \in \{1,\dots,l\}$ for any $a \in \{a_1,\dots,a_n\}$, (3) $\text{cl}'(a) \in \{l+1,\dots,m\}$ for any $a \in \{a_{n+1},\dots,a_{n+n'}\}$.
With this $\text{cl}'$, we have that $\text{cl}'(a_i) \neq \text{cl}'(a_j)$ for any $i \in \{1,\dots,n\}$ and $j \in \{n+1,\dots,n+n'\}$.
Therefore, if two points $a_i,a_j \in S$ form an inter-color dominance in with respect to $\text{cl}$, then they also form an inter-color dominance with respect to $\text{cl}'$, and vice versa.
Since the dummy points in $A$ can never contribute inter-color dominances, we have $\varGamma_{\mathcal{S}'} = \varGamma_\mathcal{S}$, which completes the proof.

\subsection{Proof of Lemma~\ref{lem-Gstar}}
Fixing $p,p' \in \{0,\dots,n\}$, we denote by $\mathcal{I}$ the collection of the independent sets $I$ of $G$ such that $|I \cap V| = p$, $|I \cap V'| = p'$.
Also, we denote by $\mathcal{I}^*$ the collection of the independent sets $I^*$ of $G^*$ such that $|I^* \cap V| = p$, $|I^* \cap V'| = p'$, $|I^* \cap U| = 3 \lambda p$, $|I^* \cap U'| = 3 \lambda n-3 \lambda p$.
It suffices to establish an one-to-one correspondence between $\mathcal{I}$ and $\mathcal{I}^*$.
Let $I \in \mathcal{I}$ be an element.
If $e = (v,v') \in E$ is an edge of $G$ (where $v \in V$ and $v' \in V'$), we say $e$ is of Type-1 if $v \in I$ (and hence $v' \notin I$), otherwise of Type-2.
Recall that for each $e \in E$, $U_e$ (resp., $U_e'$) denotes the set of the $\lambda$ vertices in $U$ (resp., $U'$) which are inserted to the edge $e$.
Now let $I^*$ be the set consists of the vertices in $I$, the vertices in $U_e$ for all Type-1 edges $e$, and the vertices in $U_e'$ for all Type-2 edges $e$.
Clearly, $I^*$ is an independent set of $G^*$.
Furthermore, by the definition of $\mathcal{I}$ and the fact that $G$ is 3-regular, we know that $G$ has $3p$ Type-1 edges and $3n-3p$ Type-2 edges.
It follows that $|I^* \cap V| = p$, $|I^* \cap V'| = p'$, $|I^* \cap U| = 3 \lambda p$, $|I^* \cap U'| = 3 \lambda n-3 \lambda p$.
Thus, $I^* \in \mathcal{I}^*$.
By mapping $I$ to $I^*$, we obtain a map from $\mathcal{I}$ to $\mathcal{I}^*$, which is obviously injective.
To see it is surjective, let $I^* \in \mathcal{I}^*$ be an element.
Set $I = I^* \cap (V \cup V')$.
We claim that $I \in \mathcal{I}$ and $I$ is mapped to $I^*$ by our map defined above.
First, since $I^*$ is an independent set of $G^*$, we must have $|I^* \cap (U_e \cup U_e')| \leq \lambda$ for any edge $e = (v,v') \in E$ of $G$ (with equality only if at least one of $v$ and $v'$ is in $I$).
But $|I^* \cap (U \cup U')| = 3 \lambda n = \lambda |E|$, which implies $|I^* \cap (U_e \cup U_e')| = \lambda$ for all $e \in E$.
It follows that for every edge $e = (v,v') \in E$, $v$ and $v'$ are not included in $I$ simultaneously, i.e., $I$ is an independent set of $G$.
In addition, $|I \cap V| = |I^* \cap V| = p$, $|I \cap V'| = |I^* \cap V'| = p'$.
Therefore, $I \in \mathcal{I}$.
To see $I$ is mapped to $I^*$, we apply again the fact that $|I^* \cap (U_e \cup U_e')| = \lambda$ for any $e \in E$.
Based on this, we further observe that for any $e \in E$, either $U_e \subseteq I^*$ or $U_e' \subseteq I^*$ (since $I^*$ is an independent set of $G^*$).
As before, we say an edge $e = (v,v') \in E$ (with $v \in V$ and $v' \in V'$) is of Type-1 if $v \in I$, otherwise of Type-2.
Note that if an edge $e \in E$ is of Type-1, we must have $U_e \subseteq I^*$ (and then $I^* \cap U_e' = \emptyset$).
Since $G$ has $3p$ Type-1 edges, $|I^* \cap U| \geq 3 \lambda p$.
But in fact $|I^* \cap U| = 3 \lambda p$ as $I^* \in \mathcal{I}^*$.
So the only possibility is that $U_e \subseteq I^*$ (and $I^* \cap U_e' = \emptyset$) for all Type-1 edges $e$ and $U_e' \subseteq I^*$ (and $I^* \cap U_e = \emptyset$) for all Type-2 edges $e$.
As a result, $I$ is mapped to $I^*$ and $|\mathcal{I}| = |\mathcal{I}^*|$, completing the proof.

%\subsection{Proof of Lemma~\ref{lem-prop}}

\subsection{Proof of Theorem~\ref{thm-embedding}}
\begin{figure}[b]
    \centering
    \includegraphics[height=3.2cm]{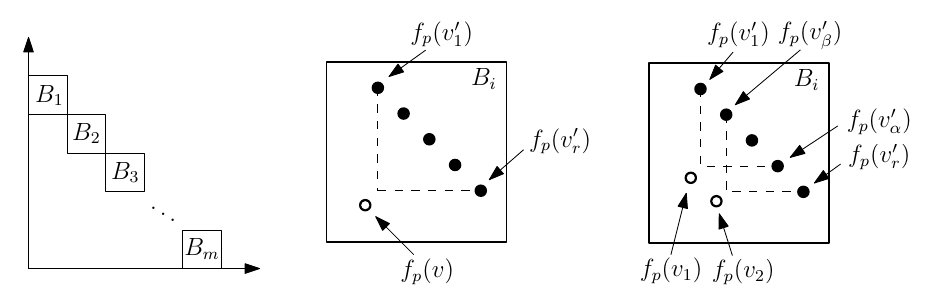}
    \caption{A local structure of $f_p$ in the box $B_i$.}
    \label{fig-Ad}
\end{figure}

Suppose $n = |V \cup V'|$.
Let $h:V \rightarrow \{1,\dots,k\}$ be a semi-discrete $k$-halfcoloring of $G$ (on $V$).
We show $\dim(G) \leq 2k$ by explicitly constructing a DPE $f:V \cup V' \rightarrow \mathbb{R}^{2k}$ of $G$.
For $i \in \{1,\dots,k\}$, we define $V_i = h^{-1}(\{i\}) \subseteq V$ (i.e., $V_i$ consists of the vertices in $V$ colored with color $i$ by $h$) and define $G_i$ as the subgraph of $G$ with the vertex set $V_i \cup V'$.
We first construct $k$ functions $f_1,\dots,f_k:V \cup V' \rightarrow \mathbb{R}^2$, and then obtain the DPE $f$ by identifying $\mathbb{R}^{2k}$ with $(\mathbb{R}^2)^k$ and ``combining'' the functions $f_1,\dots,f_k$, i.e., setting
\begin{equation*}
    f(v) = (f_1(v),\dots,f_k(v))
\end{equation*}
for all $v \in V \cup V'$.
Fixing $p \in \{1,\dots,k\}$, we describe the construction of $f_p$.
Suppose the graph $G_p$ consists of $m$ connected components.
For each $i \in \{1,\dots,m\}$, let $C_i$ be the set of the vertices in the $i$-th connected component of $G_p$.
Also, for each $i \in \{1,\dots,m\}$, let
\begin{equation*}
    B_i = \{(x,y) \in \mathbb{R}^2: i-1<x<i, m-i<y<m-i+1\}
\end{equation*}
be an open box in $\mathbb{R}^2$ (see the left part of Figure~\ref{fig-Ad}).
The function $f_p$ to be constructed maps the vertices in $C_i$ to points in $B_i$ as follows.
Since $h$ is semi-discrete, we know that $|C_i \cap V| \leq 2$.
If $|C_i \cap V| = 0$, then $C_i$ only contains an isolated vertex $v' \in V'$, and we set $f_p(v')$ to be an arbitrary point in $B_i$.
If $|C_i \cap V| = 1$, let $v$ be the only vertex in $C_i \cap V$ and suppose $C_i \cap V' = \{v_1',\dots,v_r'\}$.
In this case, we set $f_p(v_1'),\dots,f_p(v_r')$ to be a sequence of $r$ points in $B_i$ with increasing $x$-coordinates and decreasing $y$-coordinates, and $f_p(v)$ to be an arbitrary point in $B_i$ dominated by all of $f_p(v_1'),\dots,f_p(v_r')$.
See the middle part of Figure~\ref{fig-Ad} for an intuitive illustration for this case.
If $|C_i \cap V| = 2$, let $v_1,v_2$ be the two vertices in $C_i \cap V$ and again suppose $C_i \cap V' = \{v_1',\dots,v_r'\}$.
We may assume that the vertices in $C_i \cap V'$ adjacent to $v_1$ (resp., $v_2$) are exactly $v_1',\dots,v_\alpha'$ (resp., $v_\beta',\dots,v_r'$) for some $\alpha,\beta \in \{1,\dots,r\}$ with $\alpha \geq \beta$ (if not, one can easily relabel the points to achieve this).
Again, we set $f_p(v_1'),\dots,f_p(v_r')$ to be a sequence of $r$ points in $B_i$ with increasing $x$-coordinates and decreasing $y$-coordinates.
Then we set $f_p(v_1)$ to be a point in $B_i$ which is dominated by exactly $f_p(v_1'),\dots,f_p(v_\alpha')$, and set $f_p(v_2)$ to be a point in $B_i$ which is dominated by exactly $f_p(v_\beta'),\dots,f_p(v_r')$.
Note that we can definitely find such two points, since $f_p(v_1'),\dots,f_p(v_r')$ have increasing $x$-coordinates and decreasing $y$-coordinates.
In addition, by carefully determining the locations of $f_p(v_1)$ and $f_p(v_2)$ in $B_i$, we may further require that $f_p(v_1)$ and $f_p(v_2)$ do not dominate each other.
See the right part of Figure~\ref{fig-Ad} for an intuitive illustration for this case.
After considering all $C_i$, the function $f_p$ is defined for all vertices in $V_p \cup V'$ (which is the vertex set of $G_p$).
So it suffices to define $f_p$ on $V \backslash V_p$.
For each $v \in V \backslash V_p$, we simply set $f_p(v)$ to be an arbitrary point in the box $[-N,-N+1] \times [-N,-N+1]$ for a sufficiently large integer $N > 10n$ (recall that $n = |V \cup V'|$), which completes the construction of $f_p$.
We observe that $f_p$ has the following properties. \\
(1) For any $v \in V$ and $w \in V_p$, $f_p(v) \nsucc f_p(w)$. \\
(2) For any $v' \in V'$, $f_p(v')$ is not dominated by any point in the image of $f_p$. \\
(3) For any $v \in V_p$ and $v' \in V'$, $f_p(v') \succ f_p(v)$ iff $v$ and $v'$ are adjacent in $G$. \\
%(1) For any $v_1,v_2 \in V$, $f_p(v_1)$ dominates $f_p(v_2)$ only if $v_2 \notin V_p$. \\
%(2) For any $v_1',v_2' \in V'$, $f_p(v_1')$ and $f_p(v_2')$ do not dominate each other. \\
%(3) For any $v \in V$ and $v' \in V'$, $f_p(v)$ does not dominate $f_p(v')$. \\
%(4) For any $v \in V_p$ and $v' \in V'$, $f_p(v')$ dominates $f_p(v)$ iff $v$ and $v'$ are adjacent. \\
%(5) For any $v \in V \backslash V_p$ and $v' \in V'$, $f_p(v')$ dominates $f_p(v)$. \\
We do the same thing for all $p \in \{1,\dots,k\}$ and obtain the functions $f_1,\dots,f_k$.
As mentioned before, we then define $f:V \cup V' \rightarrow \mathbb{R}^{2k}$ as $f(v) = (f_1(v),\dots,f_k(v))$.
We now prove that $f$ is a DPE of $G$.
First, for any $v \in V$, we claim that $f(v)$ does not dominate any point in the image of $f$.
Indeed, $f(v) \nsucc f(v')$ for any $v' \in V'$, since $f_1(v')$ is not dominated by any point in the image of $f_1$ by the property (2) above.
Also, $f(v) \nsucc f(w)$ for any $w \in V$, since $f_p(v) \nsucc f_p(w)$ for $p = h(w)$ by the property (1) above.
Second, for any $v' \in V'$, we have that $f(v')$ is not dominated by any point in the image of $f$, simply because $f_1(v')$ is not dominated by any point in the image of $f_1$ by the property (2) above.
Finally, consider two vertices $v \in V$ and $v' \in V'$.
We claim that $f(v') \succ f(v)$ iff $v$ and $v'$ are adjacent in $G$.
If $v$ and $v'$ are adjacent, then $f_i(v') \succ f_i(v)$ for all $i \in \{1,\dots,k\}$ by the property (3) above, and hence $f(v') \succ f(v)$.
If $v$ and $v'$ are not adjacent, then $f_p(v') \nsucc f_p(v)$ for $p = h(v)$ by the property (3) above, and hence $f(v') \nsucc f(v)$.
In sum, we have $f(v') \succ f(v)$ iff $v \in V$, $v' \in V'$, $v$ and $v'$ are adjacent in $G$.
Therefore, $f$ is a DPE of $G$ to $\mathbb{R}^{2k}$.
Clearly, $f$ can be constructed in polynomial time if the $k$-halfcoloring $h$ is provided, which completes the proof of the first part of the theorem.

Next, we prove the second part of the theorem.
Again, let $h:V \rightarrow \{1,\dots,k\}$ be a semi-discrete $k$-halfcoloring of $G$ (on $V$).
Suppose $\chi_h(v') < k$ for all $v' \in V'$.
If $k = 1$, then $\chi_h(v') = 0$ for all $v' \in V'$, which implies that $G$ has no edges and thus the statement is trivial (any constant map $f:V \cup V' \rightarrow \mathbb{R}$ is a DPE of $G$).
So assume $k \geq 2$.
We show $\dim(G) \leq 2k-1$ by explicitly constructing a DPE $f:V \cup V' \rightarrow \mathbb{R}^{2k-1}$ of $G$.
In the same way as before, we define the functions $f_1,\dots,f_k: V \cup V' \rightarrow \mathbb{R}^2$.
But we need a different way to define $f$.
To this end, we first construct $k-1$ functions $f_1',\dots,f_{k-1}': V \cup V' \rightarrow \mathbb{R}^2$ based on $f_1,\dots,f_k$ as follows.
Fixing $p \in \{1,\dots,k-1\}$, we describe the construction of $f_p'$.
For all $v \in V \backslash V_k$, we set $f_p'(v) = f_p(v)$.
For all $v \in V_k$, we set $f_p'(v) = f_k(v) - (n,n)$, that is, if $f_k(v) = (x,y) \in \mathbb{R}^2$ then $f_p'(v) = (x-n,y-n)$.
Now consider the vertices in $V'$.
If a vertex $v' \in V'$ is ``adjacent'' to the color $p$ (recall that $v'$ is said to be ``adjacent'' to the color $p$ if there exists $v \in V$ adjacent to $v'$ with $h(v)=p$), then we set $f_p'(v') = f_p(v')$, otherwise $f_p'(v') = f_k(v')-(n,n)$.
By doing this for all $p \in \{1,\dots,k-1\}$, we complete constructing $f_1',\dots,f_{k-1}'$.
However, if we ``combine'' $f_1',\dots,f_{k-1}'$, we only obtain a map $V \cup V' \rightarrow \mathbb{R}^{2k-2}$ which is not guaranteed to be a DPE.
So the last ingredient needed for defining $f$ is a function $\rho:V \cup V' \rightarrow \mathbb{R}$.
The definition of $\rho$ is quite simple.
We set $\rho(v) = 1$ for all $v \in V \backslash V_k$, and $\rho(v) = 3$ for all $v \in V_k$.
For $v' \in V'$, if $v'$ is ``adjacent'' to the color $k$ or $\chi_h(v')=0$, then we set $\rho(v') = 4$, otherwise $\rho(v') = 2$.
Finally, $f:V \cup V' \rightarrow \mathbb{R}^{2k-1}$ is defined by identifying $\mathbb{R}^{2k-1}$ with $(\mathbb{R}^2)^{k-1} \times \mathbb{R}$ and ``combining'' the functions $f_1',\dots,f_{k-1}',\rho$, i.e., setting
\begin{equation*}
    f(v) = (f_1'(v),\dots,f_{k-1}'(v),\rho(v))
\end{equation*}
for all $v \in V \cup V'$.
We need to verify that $f$ is truly a DPE of $G$ to $\mathbb{R}^{2k-1}$.

First, we show that for any $v \in V$, $f(v)$ does not dominate any point in the image of $f$.
Let $v \in V$ be a vertex.
We consider two cases, $v \in V \backslash V_k$ and $v \in V_k$.
In the case of $v \in V \backslash V_k$, we first notice that $f(v) \nsucc f(w)$ for any $w \in V_k \cup V'$, simply because $\rho(v) < \rho(w)$.
To see this $f(v) \nsucc f(w)$ for any $w \in V \backslash V_k$, set $p = h(w) \neq k$.
Then $f_p'(v) = f_p(v)$ does not dominate $f_p'(w) = f_p(w)$ by the property (1) above, and hence $f(v) \nsucc f(w)$.
In the case of $v \in V_k$, we first claim that $f(v) \nsucc f(w)$ for any $w \in V$.
If $w \notin V_k$, then by setting $p = h(w) \neq k$ we have $f_p'(v) = f_k(v) - (n,n)$ does not dominate $f_p'(w) = f_p(w)$, which implies $f(v) \nsucc f(w)$.
If $w \in V_k$, then $f_1'(v) = f_k(v) - (n,n)$ does not dominate $f_1'(w) = f_k(w) - (n,n)$ since $f_k(v) \nsucc f_k(w)$ by the property (1) above, which also implies $f(v) \nsucc f(w)$.
It suffices to show that $f(v) \nsucc f(v')$ for any $v' \in V'$.
Indeed, we have either $f_1'(v') = f_1(v')$ or $f_1'(v') = f_k(v') - (n,n)$.
In each case, $f_1'(v) = f_k(v) - (n,n)$ does not dominate $f_1'(v')$ (the former case is obvious and the latter case follows from the property (2) above).
Thus $f(v) \nsucc f(v')$.

Second, we show that for any $v' \in V'$, $f(v')$ is not dominated by any point in the image of $f$.
Let $v' \in V'$ be a vertex.
By the argument above, it suffices to verify that $f(w') \nsucc f(v')$ for any $w' \in V'$.
If $v'$ is ``adjacent'' to some color $p \in \{1,\dots,k-1\}$, then we are done because $f_p'(v') = f_p(v')$ is not dominated by $f_p'(w')$ for any $w' \in V'$.
Suppose $v'$ is not ``adjacent'' to any color in $\{1,\dots,k-1\}$.
In this case, we must have $\rho(v')=4$ and $f_i'(v') = f_k(v')-(n,n)$ for all $i \in \{1,\dots,k-1\}$.
We first notice that $f(w') \nsucc f(v')$ for any $w' \in V'$ such that $\chi_h(w')>0$ and $w'$ is not ``adjacent'' to the color $k$, simply because $\rho(w') = 2 < \rho(v')$.
Then we consider the case that $w' \in V'$ is ``adjacent'' to the color $k$ or $\chi_h(w')=0$.
By the assumption $\chi_h(w')<k$, we know that $w'$ cannot be ``adjacent'' to all the $k$ colors.
In other words, if $w'$ is ``adjacent'' to the color $k$ or $\chi_h(w')=0$, $w'$ must miss some color in $\{1,\dots,k-1\}$.
Without loss of generality, we may assume $w'$ is not ``adjacent'' to the color 1.
Thus, $f_1'(v') = f_k(v')-(n,n)$ is not dominated by $f_1'(w') = f_k(w')-(n,n)$ by the property (2) above, and hence $f(w') \nsucc f(v')$.

Finally, we show that for any $v \in V$ and $v' \in V'$, $f(v') \succ f(v)$ iff $v$ and $v'$ are adjacent in $G$.
Let $v \in V$ and $v' \in V'$ be two vertices.
If $v$ and $v'$ are adjacent in $G$, one can easily verify (by checking various cases) that $\rho(v') > \rho(v)$ and $f_i'(v')$ dominates $f_i'(v)$ for all $i \in \{1,\dots,k-1\}$, which implies $f(v') \succ f(v)$.
Now suppose $v$ and $v'$ are not adjacent in $G$.
We consider two cases, $v \in V \backslash V_k$ and $v \in V_k$.
In the case of $v \in V \backslash V_k$, set $p = h(v) \neq k$.
Then $f_p'(v) = f_p(v)$.
Besides, we have either $f_p'(v') = f_p(v')$ or $f_p'(v') = f_k(v') - (n,n)$.
For the former, $f_p'(v') \nsucc f_p'(v)$ follows from the property (3) above, while for the latter $f_p'(v') \nsucc f_p'(v)$ follows obviously.
Thus, $f(v') \nsucc f(v)$.
In the case of $v \in V_k$, we have $f_i'(v) = f_k(v) - (n,n)$ for all $i \in \{1,\dots,k\}$ and $\rho(v) = 3$.
If $v'$ is not ``adjacent'' to the color $k$ and $\chi_h(v') > 0$, then $\rho(v') = 2 < \rho(v)$ and hence $f(v') \nsucc f(v)$.
If $v'$ is ``adjacent'' to the color $k$ or $\chi_h(v') = 0$, then as argued before $v'$ must miss some color in $\{1,\dots,k-1\}$.
Without loss of generality, we may assume $w'$ is not ``adjacent'' to the color 1.
Thus, $f_1'(v') = f_k(v') - (n,n)$ does not dominate $f_1'(v) = f_k(v) - (n,n)$ by the property (3) above, which implies $f(v') \nsucc f(v)$.

In sum, two vertices in $G$ share a common edge iff their images under $f$ form a dominance.
Therefore, $f$ is a DPE of $G$ to $\mathbb{R}^{2k-1}$.
It is clear that the construction of $f$ can be done in polynomial time if the $k$-halfcoloring $h$ is provided.

\subsection{Proof of Lemma~\ref{lem-halfcoloring}}
\begin{figure}[h]
    \centering
    \includegraphics[height=3cm]{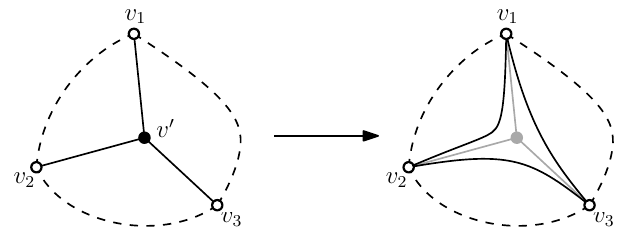}
    \caption{Deleting a vertex and adding three new edges.}
    \label{fig-Ae}
\end{figure}

Let $G=(V \cup V',E)$ be a 3-regular planar bipartite graph.
As before, we define the graph $G' = (V,E')$ by setting $E'=\{(a,b):a \sim b \text{ in }G\}$.
Then a discrete $k$-halfcoloring of $G$ on $V$ corresponds to a (conventional) $k$-coloring of $G'$ satisfying that no two adjacent vertices share the same color.
We first show that $G'$ is planar.
Fix a planar drawing $\varphi$ of $G$.
Let $v' \in V'$ be a vertex.
Since $G$ is 3-regular, $v'$ must be adjacent to three vertices $v_1,v_2,v_3 \in V$.
We now delete $v'$ as well as its three adjacent edges from $G$ and add three new edges $(v_1,v_2),(v_2,v_3),(v_3,v_1)$ to $G$.
We claim that the resulting graph is still planar.
Indeed, in the drawing $\varphi$, after we remove $\varphi(v')$ and its adjacent edges, $\varphi(v_1),\varphi(v_2),\varphi(v_3)$ will share a common face, which is the one previously containing $\varphi(v)$.
So we can draw the edges $(v_1,v_2),(v_2,v_3),(v_3,v_1)$ inside this face along with the image of the deleted edges (see Figure~\ref{fig-Ae}).
In this way, we keep deleting the vertices in $V'$ (as well as the adjacent edges) and adding new edges.
In this process, the planarity of the graph always keeps.
Until all the vertices in $V'$ are deleted, the resulting graph, which is still planar, is nothing but $G'$, as two vertices $u,v \in V$ are connected (in the resulting graph) iff $u \sim v$ in $G$.
By applying the well-known Four Color Theorem, we know that $G'$ is 4-colorable.
Furthermore, to find a 4-coloring for $G'$ can be done in quadratic time using the approach in \cite{robertson1996efficiently}.
As a result, a discrete 4-halfcoloring of $G$ can be computed in polynomial time, completing the proof.
%\newpage
\section{Computing $\varGamma_\mathcal{S}$ for \texorpdfstring{$d = 2$}{d=2} in \texorpdfstring{$O(n^2 \log^2 n)$}{O(n2 log2 n)} time}
\label{appendix:algorithm_via_2d_rangetree}
In this section, we give the details of our improved algorithm using
    2D range trees.
Formally, the 2D range tree, $\ylRT$, used in this paper is built on
    a fixed collection of planar points and maintains the weights of
    these points.
It supports the following three operations:
\begin{itemize}
    \item $\ylOpQuery(\ylRT, r)$ returns the sum of weights 
        of all the points in the query range $r$.
    \item $\ylOpUpdate(\ylRT, p, w)$ updates the weight of point $p$ to $w$.
    \item $\ylOpMul(\ylRT, r, \delta)$ multiples by a factor of
        $\delta$ the weight of every point in the range $r$.
        Note that this operation is revertible and the inverse of
        $\ylOpMul(\ylRT, r, \delta)$ is $\ylOpMul(\ylRT, r, 1/\delta)$.
\end{itemize}
With a careful implementation,
    see Appendix~\ref{appendix:rangetree:details:2016},
    all three operations can run in $O(\log^2 n)$ time.

\begin{figure}[htpb]
    \centering
    \includegraphics[]{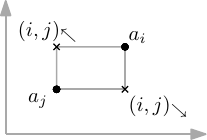}
    \caption{Illustrating $(i,j)_\searrow$ and $(i,j)_\nwarrow$ for a legal pair $(i,j)$.}
    \label{figure:algorithm_arrow}
\end{figure}

Two more notations are defined.
For a legal pair $(i, j)$,
    we use $(i, j)_\searrow$ (resp. $(i, j)_\nwarrow$) to represent
    the point $(x(a_i), y(a_j))$ (resp. $(x(a_j), y(a_i))$); see Figure~\ref{figure:algorithm_arrow}.
Also, let $\textsc{Quad}(p)$ denote the northwest open quadrant of point $p$,
    i.e., $(-\infty, x(p)) \times (y(p), \infty)$.
We now give the complete solution
    shown in Algorithm~\ref{algorithm:compute_gamma_s}
    followed by the correctness analysis.

\begin{algorithm}[h]
    \begin{algorithmic}[1] 
        \Procedure{\textsc{Compute-$\ylGammas$}}{$\mathcal{S}$}    \Comment{Recall $\mathcal{S} = (S, \text{cl}, \pi)$.}
        \State Sort all points in $S$ such that $x(a_1) < \dots < x(a_n)$.
        \State Let $\ylRT$ be the 2D range tree built on $\{(i,j)_\searrow : (i,j) \text{ is legal}\}$ with initial weights 0.
        \State Let $\ylRT_k$ be the 2D range tree built on $\{(i,j)_\searrow : (i,j) \text{ is legal and cl}(a_i)=\text{cl}(a_j)=k\}$ with initial weights 0, for every color $k$.
        \State $\textit{prod} = \prod_{i=1}^{n}{(1 - \pi(a_i))}$
        \State $\ylGammas \gets \textit{prod}$
        \State $\ylOpUpdate(\ylRT, a_0, 1)$\Comment{This implies $F(0,0)=1$. Also, no need to update $\ylRT_{\text{cl}(a_0)}$.}
        
        \For{$i \gets 1 \textbf{ to } n$} \label{algorithm:for_loop_i}
        \State $\textit{prod} \gets \textit{prod} \cdot (1 - \pi(a_i))^{-1}$
        \State $k \gets \text{cl}(a_i)$ \label{algorithm:just_after_loop_i}
        \label{algorithm:beginning_of_i_th}
        \State $\ylOpMul(
            \ylRT, 
            \textsc{Quad}(a_j),
            (1-\pi(a_j))^{-1}
        )$
        and
        $\ylOpMul(
            \ylRT_k,
            \textsc{Quad}(a_j),
            (1-\pi(a_j))^{-1}
        )$
        for all $j \in \{1, \dots, i\}$ such that
        $\text{cl}(a_j) = k$. \label{algorithm:revert1}
        \State Let $(\ell_1, \dots, \ell_i)$ be a permutation of $(1, \dots, i)$ such that $y(a_{\ell_1}) < \dots < y(a_{\ell_i})$.        
        \For{$j \gets \ell_1 \textbf{ to } \ell_i$} \label{algorithm:for_loop_j}
        \If{$(i, j)$ is a legal pair}\Comment{This implies that $\text{cl}(a_j) = k$.}
        \State $F(i,j) \gets 
        \ylOpQuery(\ylRT, \textsc{Quad}((i,j)_\nwarrow))
        -
        \ylOpQuery(\ylRT_k, \textsc{Quad}((i,j)_\nwarrow)$
        \label{algorithm:f_ij:step1}
        \State $F(i,j) \gets F(i,j) \cdot \pi^*_{i,j}$
        \label{algorithm:f_ij_step2}
        \State $\varGamma_\mathcal{S} \gets \varGamma_\mathcal{S} + F(i,j) \cdot \textit{prod}$ \label{algorithm:accumulate_gamma_s}
        \State $\ylOpMul(
            \ylRT,
            \textsc{Quad}(a_j),
            1 - \pi(a_j)
        )$
        \label{algorithm:revert2}
        \State $\ylOpMul(
            \ylRT_{k},
            \textsc{Quad}(a_j),
            1 - \pi(a_j)
        )$
        \label{algorithm:revert3}
        \EndIf
        \EndFor
        
        \State Revert all $\ylOpMul$ operations executed in Line
        \ref{algorithm:revert1},
        \ref{algorithm:revert2},
        \ref{algorithm:revert3}.
        \label{algorithm:revert_all}
        \State $\ylOpUpdate(
            \ylRT,
            (i, j)_\searrow,
            F(i, j)
        )$
        and
        $\ylOpUpdate(
            \ylRT_k,
            (i, j)_\searrow,
            F(i,j)
        )$
        for every $j \in \{1, \dots, i\}$ such that
        pair $(i, j)$ is legal.
        \label{algorithm:update_new_(i,j)}
        \State $\ylOpMul(
            \ylRT,
            (-\infty, x(a_i)) \times \mathbb{R},
            1 - \pi(a_i)
        )$
        \label{algorithm:second_last_multiple}
        \State $\ylOpMul(
            \ylRT_k,
            (-\infty, x(a_i)) \times \mathbb{R},
            1 - \pi(a_i)
        )$ 
        \EndFor
        \State $\textbf{return } \ylGammas$
        \EndProcedure
    \end{algorithmic}
    \caption{Computing $\varGamma_\mathcal{S}$ in $O(n^2 \log^2 n)$ time.}
    \label{algorithm:compute_gamma_s}
\end{algorithm}

\noindent{\bf Correctness analysis.}
We compute $F(i,j)$ for each legal pair $(i,j)$ by first enumerating $i$ from $1$ to $n$ and
	then $j$ in an order such points are visited from bottom to top;
    see the nested loop at Line~\ref{algorithm:for_loop_i} and \ref{algorithm:for_loop_j}.
For now, assume the fact, which we prove later, that the inner $j$-loop 
    correctly computes $F(i,j)$ for all legal pairs $(i, j)$ when $i$ is fixed.
We then have the following lemma.

\begin{figure}[htpb]
    \centering
    \subfloat[Points in $\parallel$.]{
        \includegraphics[width=3.2cm]{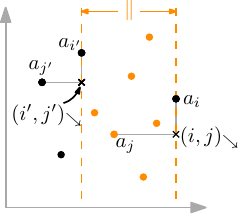}
        \label{figure:algorithm1}
    }
    \subfloat[Points in $\Box$.]{
        \includegraphics[width=3.2cm]{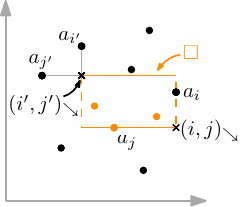}
        \label{figure:algorithm1.5}
    }
    \subfloat[Points in $\sqcap_1$.]{
        \includegraphics[width=3.2cm]{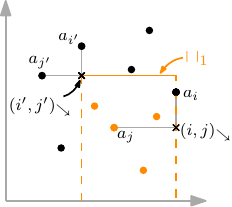}
        \label{figure:algorithm2}
    }
    \subfloat[Points in $\sqcap_2$.]{
        \includegraphics[width=3.2cm]{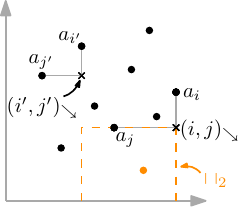}
        \label{figure:algorithm3}
    }
    \caption{Illustrating Lemma~\ref{lemma:algorithm_inner_invariant}.
        Orange color is only used to highlight each range and
        does {\em not} represent the color of each point.
        Dashed (resp. solid) boundaries are exclusive (resp. inclusive).}
\end{figure}

\begin{lemma}\label{lemma:algorithm_outer_invariant}
    At the beginning of the $i$-th iteration
	    (Line~\ref{algorithm:beginning_of_i_th}),
        the weight of $(i',j')_\searrow$ in $\ylRT$,
        such that $i' < i$,
        is equal to
        $
            F(i',j') \cdot \prod_{p \in S \cap \parallel}
            {\left(1 - \pi(p)\right)}
        $, where $\parallel$ denotes the open strip
	    $(x(a_{i'}), x(a_i)) \times \mathbb{R}$.
        (See Figure~\ref{figure:algorithm1}.)
\end{lemma}
{\em Proof.} 
This statement is trivially true for $i = 1$ as all the weights in $\ylRT$ are equal to zero except that $F(0, 0) = 1$.
Assume the statement is true for the $i$-th iteration,
    we show it also holds for the $(i+1)$-th iteration.
First, we can safely consider $\ylRT$ unchanged throughout
Line~\ref{algorithm:just_after_loop_i}-\ref{algorithm:revert_all} because
    although Line~\ref{algorithm:revert1} and \ref{algorithm:revert2} modifies
    $\ylRT$, these side-effects are reverted immediately in
    Line~\ref{algorithm:revert_all}.
After the inner $j$-loop is done, by our early assumption,
    we obtain the value of $F(i,j)$ for every legal pair $(i,j)$
    when $i$ is fixed.
These values are not currently stored in $\ylRT$ but are needed for
    the next iteration.
Thus, we update the weight of each $(i,j)_\searrow \in \ylRT$ to $F(i,j)$,
    as stated in Line~\ref{algorithm:update_new_(i,j)}.
We also need to multiply the factor $(1 - \pi(a_i))$ to the weight of
    each $(i',j')_\searrow \in \ylRT$ that is to the left of $a_i$
    because $a_i$ will be included in the strip as we proceed from $i$ to $i+1$.
This is handled by Line~\ref{algorithm:second_last_multiple}.
As such, the statement is maintained for the $(i+1)$-th iteration, which completes the proof.
\hfill $\Box$

With Lemma~\ref{lemma:algorithm_outer_invariant} in hand, we now give the proof of our
    aforementioned statement, as restated in Lemma~\ref{lemma:algorithm_inner_invariant}.
% 这里还差一句。
% 总结一下整体算法的正确性    
\begin{lemma}\label{lemma:algorithm_inner_invariant}
    Line~\ref{algorithm:f_ij:step1}-\ref{algorithm:f_ij_step2} correctly computes
    $F(i,j)$.
\end{lemma}
{\em Proof.}
Recall that 
    $
        F(i,j) =
            \pi_{i,j}^* 
            \cdot
            \sum_{(i',j') \in J_{i,j}}{F(i',j') \cdot \varPi_{i,j,i',j'}}
    $.
By Lemma~\ref{lemma:algorithm_outer_invariant}, at the beginning of the $i$-th round,
    the weight of each $(i',j')_\searrow \in \ylRT$, where $i' < i$,
    is equal to
    $
        F(i',j') \cdot \prod_{p \in S \cap \parallel}
        {\left(1 - \pi(p)\right)}
    $.
This product is off from the ideal one,
    $F(i',j') \cdot \varPi_{i,j,i',j'}$,
    by a factor of 
    $\prod_{p \in S^{(i)} \cap \Box}{(1 - \pi(p))}$, where 
    $S^{(i)} = \{ p \in S : \text{cl}(p) = \text{cl}(a_i) \}$ and 
    $\Box$ denotes the box $(x(a_{i'}), x(a_i)) \times [y(a_j), y(a_{j'})]$;
    see Figure~\ref{figure:algorithm1.5}.
To cancel this factor, we observe that
    $$
        \prod_{p \in S^{(i)} \cap \Box}{(1 - \pi(p))} =
        \prod_{p \in S^{(i)} \cap \sqcap_1}{(1 - \pi(p))} 
        \bigg/
        \prod_{p \in S^{(i)} \cap \sqcap_2}{(1 - \pi(p))},
    $$
    where 
    $\sqcap_1$ and $\sqcap_2$ respectively denote the three-sided rectangle 
    $(x(a_{i'}), x(a_i)) \times (-\infty, y(a_{j'})]$ and
    $\allowbreak(x(a_{i'}), x(a_i)) \times (-\infty, y(a_{j}))$;
    see Figure~\ref{figure:algorithm2} and \ref{figure:algorithm3}.
The former product ($\sqcap_1$) is canceled in Line~\ref{algorithm:revert1},
    and the latter ($\sqcap_2$) is gradually accumulated back via
    $(j - 1)$ calls of Line~\ref{algorithm:revert2} as
    $a_{\ell_1}, \dots, a_{\ell_{j-1}}$ are all below $a_{\ell_j}$.
Thus, the weight of each $(i',j')_\searrow \in \ylRT$ is equal to
    $F(i',j') \times \varPi_{i,j,i',j'}$ right before $F(i,j)$ gets evaluated.
Finally, the range query in Line~\ref{algorithm:f_ij:step1}
    sums up the weight of every $(i',j')_\searrow \in \ylRT$ such that
    $(i',j') \in J_{i,j}$.
(Note that the subtraction in Line~\ref{algorithm:f_ij:step1} is needed
    because $\ylOpQuery(\ylRT, \textsc{Quad}((i,j)_\nwarrow))$
    also counts the probabilities
    of those legal pairs that have the same color as $\text{cl}(a_i)$.)
Therefore, the value of $F(i,j)$ is correctly computed after
    Line~\ref{algorithm:f_ij_step2}.
\hfill $\Box$

Though Lemma~\ref{lemma:algorithm_outer_invariant} and \ref{lemma:algorithm_inner_invariant}
    are cross-referencing, one can easily figure out that
    this is not a circular reasoning and is indeed a valid proof.
Also, both lemmas can directly apply to $\ylRT_k$'s as we always 
    query/update $\ylRT$ and $\ylRT_k$'s in the same way.
Finally, all $F(i,j)$'s are computed and added up into
    $\varGamma_\mathcal{S}$,
    which completes the correctness proof of the entire algorithm.

The overall runtime of Algorithm~\ref{algorithm:compute_gamma_s}
    is $O(n^2 \log^2 n)$ since there are $O(n^2)$ range queries/updates, 
    each of which takes $O(\log^2 n)$ time.
The space occupied by $\ylRT$, denoted by $|\ylRT|$, is $O(n^2 \log n^2) = O(n^2 \log n)$ as
    there are $O(n^2)$ legal pairs.
Similarly, let $n_k$ be the number of points in color $k$, and then
    $\ylRT_k$ costs $O(n_k^2 \log n_k)$ space.
Assume there are $K$ colors in total.
We have $n_1 + \dots + n_K = n$ and thus
    $|\ylRT_1| + \dots + |\ylRT_K| = O(n^2 \log n)$.
The overall space complexity is $O(|\ylRT| + |\ylRT_1| + \dots + |\ylRT_K|) =
    O(n^2 \log n)$.

\subsection{Implementation details of our 2D range tree}
\label{appendix:rangetree:details:2016}
In this section, we discuss how to implement an augmented 2D range tree,
    $\ylRT$, to dynamically support
    $\ylOpQuery, \ylOpUpdate$, and $\ylOpMul$ in $O(\log^2 m)$ time,
    where $m$ is the input size.

We first describe how to implement a dynamic 1D range tree, $\ylRTOneD$,
    built on the $y$-coordinates of a set of planar points, $P$, to support
    the three operations, where the range used in $\ylOpQuery$ and 
    $\ylOpMul$ is a 1D interval.
The leaves, sorted by increasing $y$-coordinates, of $\ylRTOneD$
    are points in $P$ with initial weight equal to 0.
In addition, in each internal node, $u$, we store two fields, 
    $\mathit{sum}(u)$ and $\mathit{mul}(u)$, where the former is the sum
    of weights in the subtree rooted at $u$ and the latter is the
    multiplication-factor that needs to be applied to all the nodes
    in the subtree.
For simplicity we use the notion $\mathit{sum}(u)$ to denote the weight
    of $u$ if it is a leaf.
Also, set $sum(u) = 0$ and $mul(u) = 1$ initially.

Given a query/update range, we first identify $O(\log m)$
    canonical nodes, $\mathcal{C}$, of $\ylRTOneD$ via a recursive
    down-phase traversal.
We then aggregate or modify the data in each canonical node.
Finally, we refine the fields of those nodes along the path from every
    canonical node up to the root, as the recursion gradually terminates.

In the down-phase, when a non-leaf node $u$ is visited, we call the 
    following $\textsc{Push}$ method to revise $sum(u)$ based on $mul(u)$ 
    and then push the factor further to its two children.
In the up-phase, we apply the $\textsc{Combine}$ method to each node to 
    readjust the sum.
Between the down and up phase, we perform one of the following three operations.
\begin{itemize}
    \item Add up $\mathit{sum}(u)$ for every $u \in \mathcal{C}$ for 
        $\ylOpQuery(\ylRTOneD, p)$.
    \item Update $\mathit{sum}(u)$ to $w$ for the {\em only} element
        $u \in \mathcal{C}$ for $\ylOpUpdate(\ylRTOneD, p, w)$.
    \item Multiply $\mathit{mul}(u)$ by a factor of $\delta$ for every
        $u \in \mathcal{C}$ for $\ylOpMul(\ylRTOneD, p, \delta)$.
\end{itemize}

\begin{algorithm}[h]
    \begin{algorithmic}[1]
        \Procedure{\textsc{Push}}{$u$} \Comment{Only called in the down-phase.}
            \State $\mathit{sum}(u) \leftarrow \mathit{sum}(u) \cdot \mathit{mul}(u)$
            \If{$u$ is not a leaf}
                \State $\mathit{mul}(\mathit{lchild}(u)) \leftarrow \mathit{mul}(\mathit{lchild}(u)) \cdot mul(u)$
                \State $\mathit{mul}(\mathit{rchild}(u)) \leftarrow \mathit{mul}(\mathit{rchild}(u)) \cdot mul(u)$
            \EndIf
            \State $\mathit{mul} \leftarrow 1$
        \EndProcedure
        
        \Procedure{\textsc{Combine}}{$u$} \Comment{Only called in the up-phase and we must have $\mathit{mul}(u) = 1$.}
            \State $\mathit{sum(u)} \leftarrow \mathit{sum}(\mathit{lchild}(u)) + \mathit{sum}(\mathit{rchild}(u))$
        \EndProcedure
    \end{algorithmic}
    \caption{Implementation details of $\textsc{Push}$ and $\textsc{Combine}$.}
\end{algorithm}

Finally, we build our 2D range tree, $\ylRT$,
    on the $x$-coordinates of the given input.
For each node $u \in \ylRT$, we build an aforementioned 1D range tree
    w.r.t. the set of points in $u$. 
We also store at $u$ a similar tag indicating the multiplication factor
    that needs to be applied to the 1D range tree stored at $u$ as well as
    all $u$'s descendants.
Given a 2D range query, we do a down-phase traversal identifying 
    $O(\log m)$ canonical nodes of $\ylRT$.
For each visited node $u$ during the traversal,
    we should apply the multiplication tag to the 1D tree stored at $u$ and
    push it further to $u$'s two children.
This takes $O(\log m)$ time.
Then, for every canonical node $u$, we spend another
    $O(\log m)$ time querying the 1D range tree stored at $u$,
    as stated above.
Therefore, all three operations can be done in $O(\log^2 m)$ time, and 
    $\ylRT$ occupies $O(m \log m)$ space.

\subsection{Handling range-multiplication/division with a factor of zero}
One may notice that the implementation above contains a flaw for 
    $\ylOpMul(\ylRT, r, \delta)$ when $\delta = 0$ because
    the inverse of this operation does not exist as $1 / 0$ is undefined.
We can overcome this issue by adding in each node a {\em zero-counter} and 
    counting the number of zero factors separately.
That is, if $\ylOpMul$ multiplies a factor of zero, 
    we increment the zero-counter of each canonical node instead of 
    modifying $\mathit{sum}$ and $\mathit{mul}$ fields;
    if $\ylOpMul$ divides a factor of zero, we decrement the corresponding
    zero-counters.
Also, when a $\ylOpQuery$ is triggered,
    we simply return zero for those canonical nodes whose zero-counter
    is positive.
This solves the problem without increasing the runtime of all three operations.
    
%\newpage
\section{A generalization of Lemma~\ref{lem-CSISC}} \label{app-general}
In this section, we extend Lemma~\ref{lem-CSISC} to a general result revealing the hardness of stochastic geometric problems (under existential uncertainty).
Many stochastic geometric problems focus on computing the probability that a random sample of the given stochastic dataset has some specific property, e.g., \cite{agarwal2014convex,fink2016hyperplane,kamousi2014closest,xue2016separability} and this paper.
This kind of problems can be abstracted and generalized as follows.
Let $\mathcal{C}$ be a category of geometric objects (e.g., points, lines, etc.), and $\mathbf{P}$ be a property defined on finite sets of objects in $\mathcal{C}$.
\begin{definition}
We define the $\mathbf{P}$-\textit{probability-computing} problem as follows.
The input is a stochastic dataset $\mathcal{S} = (S,\pi)$ where $S$ is a set of objects in $\mathcal{C}$ and $\pi:S \rightarrow (0,1]$ is the function defining existence probabilities for the objects.
The goal is to compute the probability that a random sample of $\mathcal{S}$ has the property $\mathbf{P}$.
\end{definition}
\textbf{Example 1.} Let $\mathcal{C}$ be the category of points in $\mathbb{R}^d$, and $\mathbf{P}$ be the property that the convex-hull of the set of points (in $\mathbb{R}^d$) contains a fixed point $q \in \mathbb{R}^d$.
In this case, the $\mathbf{P}$-probability-computing problem is the convex-hull membership probability problem \cite{agarwal2014convex}.
\medskip

\noindent
\textbf{Example 2.} Let $\mathcal{C}$ be the category of bichromatic points in $\mathbb{R}^d$, and $\mathbf{P}$ be the property that the set of bichromatic points is linearly separable.
In this case, the $\mathbf{P}$-probability-computing problem is the stochastic linear separability problem \cite{fink2016hyperplane,xue2016separability}.
\medskip

\noindent
\textbf{Example 3.} Let $\mathcal{C}$ be the category of points in $\mathbb{R}^d$, and $\mathbf{P}$ be the property that the closest-pair distance of the set of points is at most a fixed threshold $\ell$.
In this case, the $\mathbf{P}$-probability-computing problem is the stochastic closest-pair problem \cite{kamousi2014closest}.
\medskip

By generalizing Definition~\ref{def-CSISC}, we may also consider an abstract notion of the cardinality-sensitive-counting problem.
\begin{definition}
Let $c$ be a fixed constant.
We define the $\mathbf{P}$-\textit{cardinality-sensitive-counting} problem as follows.
The input consists of a set $S$ of objects in $\mathcal{C}$ and a $c$-tuple $(S_1,\dots,S_c)$ of disjoint subsets of $S$.
The goal is to compute, for every $c$-tuple $(n_1,\dots,n_c)$ of integers where $0 \leq n_i \leq |S_i|$, the number of the subsets $S' \subseteq S$ which have the property $\mathbf{P}$ and satisfy $|S' \cap S_i|=n_i$ for all $i \in \{1,\dots,c\}$.
\end{definition}
\textbf{Example 4.} Consider the following problem: given a set $S$ of bichromatic (red/blue) points in $\mathbb{R}^d$, compute the number of the linearly separable subsets $S' \subseteq S$ which contain an equal number of red and blue points.
This is clearly a restricted version of the $\mathbf{P}$-cardinality-sensitive-counting problem, where $\mathcal{C}$ is the category of bichromatic points in $\mathbb{R}^d$ and $\mathbf{P}$ is the property that the set of bichromatic points is linearly separable.
\medskip

The following theorem, which generalizes Lemma~\ref{lem-CSISC}, implies that the $\mathbf{P}$-probability-computing problem is at least as ``hard'' as the $\mathbf{P}$-cardinality-sensitive-counting problem.
The proof is (almost) the same as that of Lemma~\ref{lem-CSISC}.
\begin{theorem}
The $\mathbf{P}$-cardinality-sensitive-counting problem is polynomial-time reducible to the $\mathbf{P}$-probability-computing problem for any $\mathbf{P}$.
\end{theorem}

%\newpage
\section{Implication in order dimension theory} \label{app-order}
In Section~\ref{sec-d>=7}, we achieved the result that $\dim(G) \leq 7$ for any 3-regular planar bipartite graph $G$.
In this section, we establish an implication of this result in order dimension theory \cite{trotter2001combinatorics}.
Let $X$ be a finite set and $<_P$ be a partial order on $X$.
A set $\{<_1,\dots,<_t\}$ of linear orders (or total orders) on $X$ is said to be a \textit{realizer} of $<_P$ if
\begin{equation*}
    <_P = \bigcap_{i=1}^{t} <_i,
\end{equation*}
that is, for any $x,y \in X$, $x <_P y$ iff $x <_i y$ for all $i \in \{1,\dots,t\}$.
The \textit{order dimension} $\dim(<_P)$ of $<_P$ is defined as the least cardinality of a realizer of $<_P$ (see for example \cite{trotter2001combinatorics}).

The partial order $<_P$ can be represented by a transitive directed graph $G_{<_P} = (X,E)$ where $E = \{\langle x,y \rangle: x <_P y\}$.
The \textit{comparability graph} $H_{<_P}$ of $<_P$ is defined as the underlying undirected graph of $G_{<_P}$, i.e., $H_{<_P} = (X,E')$ where $E' = \{(x,y) : x <_P y\}$.
Our result implies the following.
\begin{corollary}
Let $(X,<_P)$ be a partial ordered set.
If the comparability graph $H_{<_P}$ is 3-regular planar bipartite, then $\dim(<_P) \leq 7$.
\end{corollary}
\textit{Proof.}
Suppose $H_{<_P} = (X_1 \cup X_2,E)$, which is 3-regular planar bipartite.
We must construct a realizer of $<_P$ of size at most 7.
Without loss of generality, we may assume $H_{<_P}$ is connected (otherwise we could work on each connected components separately).
%Consider the transitive directed graph $G_{<_P}$ defined above.
%Since $H_{<_P}$ is connected, the edges in $G_{<_P}$ must be all directed from $X_1$ to $X_2$ or all directed from $X_2$ to $X_1$ (otherwise $G_{<_P}$ is not transitive), assume the former.
Using our result in Section~\ref{sec-d>=7}, we have $\dim(H_{<_P}) \leq 7$, so there exists a DPE $f:X_1 \cup X_2 \rightarrow \mathbb{R}^7$ of $H_{<_P}$.
Let $Y$ be the image of $f$.
By Lemma~\ref{lem-regular}, we may further assume that $Y$ is regular in $\mathbb{R}^7$.
Now define a directed graph $G_f = (X,E_f)$ as $E_f = \{\langle x,y \rangle: f(y) \succ f(x)\}$.
It is clear that $G_f$ is transitive and $H_{<_P}$ is the underlying undirected graph of $G_f$.
Since $H_{<_P}$ is connected, the edges in $G_f$ must be all directed from $X_1$ to $X_2$ or all directed from $X_2$ to $X_1$ (otherwise $G_f$ is not transitive).
On the other hand, $H_{<_P}$ is also the underlying undirected graph of $G_{<_P}$ (defined above) and $G_{<_P}$ is also transitive.
Therefore, either $G_f = G_{<_P}$ or $G_f$ and $G_{<_P}$ are ``reverses'' of each other ($G_f$ is the same as $G_{<_P}$ except the orientations of the edges are reversed).
If $G_f = G_{<_P}$, we define a set $\{<_1,\dots,<_7\}$ of linear orders on $X$ as $x <_i y$ iff the $i$-th coordinate of $f(x)$ is smaller than $i$-th coordinate of $f(y)$.
If $G_f$ and $G_{<_P}$ are reverses of each other, we define $\{<_1,\dots,<_7\}$ as $x <_i y$ iff the $i$-th coordinate of $f(x)$ is greater than $i$-th coordinate of $f(y)$.
Since $Y$ is regular, $<_1,\dots,<_7$ are truly linear orders on $X$.
It suffices to verify that $<_P = \bigcap_{i=1}^{7} <_i$.
We only verify for the case of $G_f = G_{<_P}$, the other case is similar.
Suppose $x <_P y$.
Then $\langle x,y \rangle$ is an edge of $G_{<_P}$ and also an edge of $G_f$.
By the definition of $G_f$, we have $f(y) \succ f(x)$, which implies $x <_i y$ for all $i \in \{1,\dots,7\}$.
Suppose $x <_i y$ for all $i \in \{1,\dots,7\}$.
Then $f(y) \succ f(x)$.
Hence, $\langle x,y \rangle$ is an edge of $G_f$ and also an edge of $G_{<_P}$.
It follows that $x <_P y$.
\hfill $\Box$

\end{document}